\documentstyle[preprint,aps,floats,psfig]{revtex}
\newcommand {\baralpha}	{\overline{\alpha}}

\newcommand {\barssq}	{\overline{s}^2}

\newcommand {\bargzsq}	{\overline{g}_Z^2}

\newcommand {\baralphaqsq}	{\overline{\alpha}(q^2)} 
\newcommand {\baresqqsq} 	{\overline{e}^2(q^2)}

\newcommand {\barssqqsq}	{\overline{s}^2(q^2)}

\newcommand {\bargzsqqsq}	{\overline{g}_Z^2(q^2)}
\newcommand {\bargwsqqsq}	{\overline{g}_W^2(q^2)}

\newcommand {\qsq} {q^2}

\newcommand {\barf}      {\overline{f}}
\newcommand {\barb}      {\overline{b}}
\newcommand {\bardeltab} {\overline{\delta}_b}

\newcommand {\hate}	{\hat{e}}
\newcommand {\hatg}	{\hat{g}}
\newcommand {\hats}	{\hat{s}}
\newcommand {\hatc}	{\hat{c}}
\newcommand {\hatgz}	{\hat{g}_Z}
 
\newcommand {\hatalpha}	{\hat{\alpha}} 
\newcommand {\hatesq}	{\hat{e}^2}
\newcommand {\hatgsq}	{\hat{g}^2}
\newcommand {\hatssq}	{\hat{s}^2}
\newcommand {\hatcsq}	{\hat{c}^2}
\newcommand {\hatgzsq}	{\hat{g}_Z^2}

\newcommand {\mwsq}  {m_W^2}
\newcommand {\mzsq}  {m_Z^2}
\newcommand {\Lamsq} {\Lambda^2}
\newcommand {\mhsq}  {m_H^2}

\newcommand {\fdw}     {f_{DW}} 
\newcommand {\fdb}     {f_{DB}}
\newcommand {\fbw}     {f_{BW}}
\newcommand {\fpone}   {f_{\Phi,1}}
\newcommand {\fwww}    {f_{WWW}}
\newcommand {\fw}      {f_{W}}
\newcommand {\fb}      {f_{B}}
\newcommand {\fww}     {f_{WW}}
\newcommand {\fbb}     {f_{BB}}

\newcommand {\odw}     {{\cal O}_{DW}} 
\newcommand {\odb}     {{\cal O}_{DB}}
\newcommand {\obw}     {{\cal O}_{BW}}
\newcommand {\opone}   {{\cal O}_{\Phi,1}}
\newcommand {\owww}    {{\cal O}_{WWW}}
\newcommand {\ow}      {{\cal O}_{W}}
\newcommand {\ob}      {{\cal O}_{B}}
\newcommand {\oww}     {{\cal O}_{WW}}
\newcommand {\obb}     {{\cal O}_{BB}}
\newcommand {\optwo}   {{\cal O}_{\Phi,2}}
\newcommand {\opthree} {{\cal O}_{\Phi,3}}
\newcommand {\opfour}  {{\cal O}_{\Phi,4}}

\newcommand {\pitgg}    {\Pi_T^{\gamma\gamma}}
\newcommand {\pitgz}    {\Pi_T^{\gamma Z}}
\newcommand {\pitzz}    {\Pi_T^{ZZ}}
\newcommand {\pitww}    {\Pi_T^{WW}}
\newcommand {\barpitgg} {\overline{\Pi}_T^{\gamma\gamma}}
\newcommand {\barpitgz} {\overline{\Pi}_T^{\gamma Z}}
\newcommand {\barpitzz} {\overline{\Pi}_T^{ZZ}}
\newcommand {\barpitww} {\overline{\Pi}_T^{WW}}

\newcommand {\barpitggg} {\overline{\Pi}_{T,\gamma}^{\gamma\gamma}}
\newcommand {\barpitggz} {\overline{\Pi}_{T,\gamma}^{\gamma Z}}

\newcommand {\barpitzzz} {\overline{\Pi}_{T,Z}^{ZZ}}
\newcommand {\barpitwww} {\overline{\Pi}_{T,W}^{WW}}

\newcommand {\rgg} {R_{\gamma\gamma}}
\newcommand {\rgz} {R_{\gamma Z}}
\newcommand {\rzz} {R_{ZZ}}
\newcommand {\rww} {R_{WW}}

\newcommand {\oetwo}   {${\cal O}(E^2)\,$}
\newcommand {\oefour}  {${\cal O}(E^4)\,$}
\newcommand {\oesix}   {${\cal O}(E^6)\,$}
\newcommand {\oeeight} {${\cal O}(E^8)\,$}

\newcommand {\call}  {{\cal L}}
\newcommand {\trace} {{\rm Tr}} 

\newcommand {\kapgam}  {\kappa_\gamma}
\newcommand {\kapz}    {\kappa_Z}
\newcommand {\lamgam}  {\lambda_\gamma}
\newcommand {\lamz}    {\lambda_Z}

\newcommand {\gonez}   {g_1^Z}

\newcommand {\loglammu} {\log\bigg(\frac{\Lambda^2}{\mu^2}\bigg)}
\newcommand {\loglamhatmu} {\log\bigg(\frac{\Lambda^2}{\hat{\mu}^2}\bigg)}
\newcommand {\loglammh} {\log\bigg(\frac{\Lambda^2}{m_H^2}\bigg)}
\newcommand {\loglammz} {\log\bigg(\frac{\Lambda^2}{m_Z^2}\bigg)}

\draft

\preprint{\vbox{\baselineskip14pt
\hbox{\bf KEK-TH-519}
\hbox{\bf KEK Preprint 97-88}
\hbox{\bf BNL-HET-SD-97-003}
\hbox{\bf hep-ph/9706542}
\hbox{June 1997}}}
\title{Low-Energy Constraints on New Physics Revisited}

\author{S.~Alam$^{1,2}$, S.~Dawson$^3$ and R.~Szalapski$^1$}
\address{$^1$Theory Group, KEK, Tsukuba, Ibaraki 305, Japan\\
$^2$Physics Department, University of Peshawar, Peshawar, NWFP, Pakistan\\
$^3$Physics Department, Brookhaven National Laboratory, Upton, NY 
11973, USA}

\begin{document}
\maketitle 


\begin{abstract}
It is possible to place constraints on non-Standard-Model gauge-boson 
self-couplings and other new physics by studying their one-loop contributions 
to precisely measured 
observables.  We extend previous analyses which constrain such nonstandard 
couplings, and we present the results in a compact and transparent form.  
Particular attention is given to comparing results for the light-Higgs scenario, 
where nonstandard effects are parameterized by an effective Lagrangian with a 
linear realization of the electroweak symmetry breaking sector, and the 
heavy-Higgs/strongly interacting scenario, described by the electroweak chiral 
Lagrangian.  The constraints on nonstandard gauge-boson self-couplings which are
obtained from a global analysis of low-energy data and LEP/SLC measurements on 
the $Z$ pole are updated and improved from previous studies.  
\end{abstract}
\newpage
 
\section{Introduction}\label{sec-intro}

Due to the extraordinary precision of electroweak data at low energy and on the 
$Z$ pole it is possible to place constraints on models for physics beyond the 
Standard Model (SM) by studying the loop-level contributions of the new physics
to electroweak observables.  Gauge-boson self-interactions are a fascinating 
aspect of the SM, and the exploration of this sector is still in its early 
stages.  While this sector is important in its own right, it is intimately
related to the symmetry-breaking sector of the SM.  Hence, we are strongly
motivated to garner from the body of electroweak precision data any and 
all available clues concerning these heretofore more poorly understood sectors
of the SM.

Currently all available precision data concerns processes with four external
light fermions (such as $e^+e^-\rightarrow Z^*\rightarrow
f\barf$ at LEP).
  We follow the scheme of Ref.~\cite{hhkm94} which organizes the 
calculation of these amplitudes in the following manner.  First we calculate 
$\pitgg(\qsq)$, $\pitgz(\qsq)$, $\pitzz(\qsq)$ and 
$\pitww(\qsq)$, {\em i.e.} the transverse components of the $\gamma\gamma$, 
$\gamma Z$, $ZZ$ and $WW$ two-point-functions, respectively.
As well we must calculate $\Gamma^{ff\gamma}(\qsq)$, $\Gamma^{ffZ}(\qsq)$
and $\Gamma^{ff^\prime W}(\qsq)$, {\em i.e.} corrections to the 
gauge-boson-fermion vertices.  The two-point-functions and a portion of the 
vertex corrections are combined {\em via} the pinch 
technique\cite{kl89,cp89andpap90,ds92anddks93,pinch} to form the 
gauge-invariant effective charges, $\baresqqsq$, $\barssqqsq$, 
$\bargzsqqsq$ and $\bargwsqqsq$.  These effective charges 
contain the major part of the higher order corrections and are universal 
to all four-fermion amplitudes.  (Hence, this approach is especially well 
suited to a global analysis of electroweak precision data.)  The calculation of 
the four-fermion amplitudes is then completed by adding the process-dependent 
vertex and box corrections.  A more complete discussion is given in 
Section~\ref{sec-charges}.  In fact, most of the technical details are provided
in Section~\ref{sec-charges}, which allows us to be very much to the point in 
the ensuing sections.

In the context of this paper all of the non-SM contributions enter {\em via} the 
effective charges plus a form factor for the $Zb\barb$ vertex.  
With the exception of this latter form factor, the vertex and box corrections 
reduce to their SM values for the quantities we compute.  This greatly 
simplifies the analysis.

In Section~\ref{sec-linear} the SM Lagrangian is extended by the addition of 
energy-dimension-six (\oesix) operators.  The operators are constructed 
from the fields of the low-energy spectrum including the usual SU(2) Higgs 
doublet of the SM, {\em i.e.} spontaneous symmetry breaking (SSB) is linearly 
realized.  The effective charges and the $Zbb$-vertex form factor, 
$\bardeltab$\cite{hhkm94}, are calculated in this scheme.  In 
Section~\ref{sec-nonlinear} the electroweak chiral Lagrangian, in which there is 
no physical Higgs boson and the symmetry breaking is nonlinearly 
realized\cite{ab80}, is discussed, and we repeat the calculation of the 
effective charges and $\bardeltab$.  Then, in Section~\ref{sec-nabgbv}, 
we specialize to a discussion of non-Abelian gauge boson couplings.

Numerical results are given in Section~\ref{sec-numerical}. We pay particular 
attention to the uncertainties inherent in obtaining bounds  on new physics
from one-loop effects.   First, the sensitivity of the data to the parameters of 
the effective Lagrangians of Section~\ref{sec-linear} and 
Section~\ref{sec-nonlinear} is estimated by considering the contributions of 
only one new operator at a time.  Then, bounds on non-SM contributions to 
gauge-boson self-couplings are presented accounting for limited correlations.  
Additionally we consider some more complicated scenarios, and we compare the 
results from both the linear and the nonlinear models.  

\section{Low-energy parameters and effective charges}\label{sec-charges}

\begin{figure}[tb]
\begin{center}
\leavevmode\psfig{file=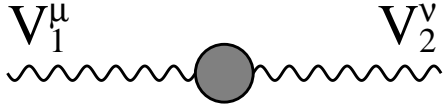,angle=0,height=1.5cm,silent=0}
\vspace*{.7in} 
\end{center}
\caption{Higher-order contributions to the $V_1^\mu V_2^\nu$
 two-point-functions;
$V_1V_2$ denotes $\gamma\gamma$, $\gamma Z$, $ZZ$ or $WW$.  Generically the 
`blob' may represent a contact term, a `bubble' or a `tadpole'.}
\label{fig-pi} 
\end{figure}

We begin by calculating the corrections to the gauge-boson two-point-functions 
as depicted by Fig.~\ref{fig-pi}.  Introducing the transverse and longitudinal 
\vspace{.5in} 
projection operators
\begin{equation}
\label{pt-pl}
P_T^{\mu\nu} = g^{\mu\nu} - \frac{q^\mu q^\nu}{\qsq}\;,\makebox[1cm]{}
P_L^{\mu\nu} = \frac{q^\mu q^\nu}{\qsq}\;,
\end{equation}
which possess the desirable properties
\begin{equation}
\label{properties}
P_T^{\mu\nu} + P_L^{\mu\nu} = g^{\mu\nu},\makebox[0.5cm]{}
P_{T \alpha}^{\mu}P_{T}^{\alpha\nu} = P_T^{\mu\nu},\makebox[0.5cm]{}
P_{L \alpha}^{\mu}P_{L}^{\alpha\nu} = P_L^{\mu\nu},\makebox[0.5cm]{}
P_{T \alpha}^{\mu}P_{L}^{\alpha\nu} = 0 = P_{L \alpha}^{\mu}P_{T}^{\alpha\nu},
\end{equation}
we may write the result of the calculation of Fig.~\ref{fig-pi} as
\begin{equation}
\label{pimunu}
-i\Pi^{\mu\nu}_{V_1V_2}(\qsq) = -i \Pi^{V_1V_2}_T(\qsq) P_T^{\mu\nu} 
                          -i \Pi^{V_1V_2}_L(\qsq) P_L^{\mu\nu}\;,
\end{equation}
where $\qsq$ is the four-momentum squared of the propagating gauge bosons.
Since we are considering processes where the gauge-boson propagators are 
coupled to massless fermion currents, we need to consider only the transverse
contribution, $\Pi^{V_1V_2}_T(\qsq)$; the longitudinal contributions don't 
contribute by the Dirac equation for massless fermions.  Equivalently we can 
calculate $-i\Pi^{\mu\nu}_{V_1V_2}(\qsq)$ and retain only the coefficient of 
$-ig^{\mu\nu}$.

\begin{figure}[tb]
\begin{center}
\leavevmode\psfig{file=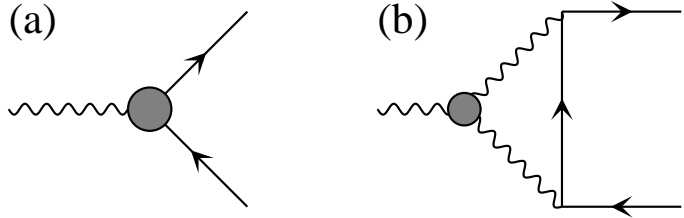,angle=0,height=3cm,silent=0}
\end{center}
\caption{Higher-order contributions to the $Vf_1f_2$ vertex where 
$V = \gamma$, $Z$ or $W$.  Generically the `blob' in (a) denotes a large variety 
of graphs.  However, for the new-physics contributions which we discuss, all
contributions arise from graphs of type (b).}
\label{fig-vertex} 
\end{figure}

Next, we calculate vertex corrections as depicted in Fig.~\ref{fig-vertex}.
Using the pinch technique, a portion of the vertex corrections in 
Fig.~\ref{fig-vertex}(a) are combined with the two-point-function corrections.
This standard technique\cite{kl89,cp89andpap90,ds92anddks93,pinch} renders
propagator and vertex corrections separately gauge invariant. Furthermore,
large cancellations which would occur between the propagator and vertex 
contributions are avoided.  

For the SM contributions we use the results of 
Ref.~\cite{hhkm94}.  For the new-physics contributions 
which we consider in 
later sections the discussion is very simple.  All new-physics contributions 
are of the type depicted in Fig.~\ref{fig-vertex}(b), where the `blob' denotes
some nonstandard contribution to the $WW\gamma$ or $WWZ$ vertex.  These 
corrections can be divided into two pieces.  One piece, which is
independent of any fermion masses, is purely pinch term; the remaining 
contributions, which depend on the fermion masses, will remain as part of 
the vertex corrections.  We will discuss these latter corrections later in 
this section.

For the moment we neglect the contributions of fermion masses, and, following 
Ref.~\cite{hisz93}, we write
\begin{mathletters}
\label{del-Gam}
\begin{eqnarray}
\label{del-Gam-g}
-i \Delta\Gamma_\mu^{\gamma f_1f_2}(q) & = & 
   -i \gamma_\mu \frac{1}{2}(1-\gamma_5) \hatg I_3^f 
   \Delta\Gamma_L^{\gamma}(\qsq) \;,\\
\label{del-Gam-z}
-i \Delta\Gamma_\mu^{Z f_1f_2}(q) & = & 
   -i \gamma_\mu \frac{1}{2}(1-\gamma_5) \hatg I_3^f 
   \Delta\Gamma_L^{Z}(\qsq)\;,\\
\label{del-Gam-w}
-i \Delta\Gamma_\mu^{W f_1f_2}(q) & = & 
   -i \gamma_\mu \frac{1}{2}(1-\gamma_5) \frac{\hatg}{\sqrt{2}} 
   \Delta\Gamma_L^{W}(\qsq)\;,
\end{eqnarray}
\end{mathletters}
where $I_3^f=\pm\frac{1}{2}$ 
is the third component of weak isospin for the external fermion.
The notation on the left-hand side should be clear from the superscripts.
Here and through the remainder of the paper we separate various quantities 
according to $X = X_{\rm SM} + \Delta X$.  Hence, above, $\Delta\Gamma$ is the 
contribution of the new physics to the vertex correction, $\Gamma$ (indices 
suppressed for brevity).  All `hatted'
couplings are the $\overline{\rm MS}$ couplings, and hence they satisfy the 
tree-level relations $\hate = \hatg\hats = \hatgz\hats\hatc$ and
$\hatesq = 4\pi\hatalpha$.  In particular, $\hatg$ is the SU(2) 
coupling, $\hats$ and $\hatc$ are the sine and cosine of the weak mixing 
angle, and the strength of the photon coupling is given by $\hate$ or 
$\hatalpha$.  Finally, the U(1) coupling is given by $\hatg^\prime 
= \hatg\hats/\hatc$.

Notice in Eqn.~(\ref{del-Gam}) that the corrections are purely 
left-handed due to the coupling of at least one $W$ boson to the fermion line,
hence we have extracted a factor of $\frac{1}{2}(1-\gamma_5)$ on the 
right-hand side.  The appearance of the factor $I_3^f$ in 
Eqns.~(\ref{del-Gam-g})-(\ref{del-Gam-z}) may be understood as follows.  
For corrections to the $WW\gamma$ or $WWZ$ vertex due to the type of loop
graph depicted in Figure~\ref{fig-vertex}(b), there are two internal $W$ bosons, 
one of each charge, connected to an external photon or $Z$ boson through a 
$WW\gamma$ or $WWZ$ vertex.  If the external fermion
legs are up-type quarks, then the internal fermion is a down-type quark (and 
{\em vice versa}).  Interchanging the up-type and down-type quarks interchanges
the $W^+$ and $W^-$, which, due to the properties of the $WW\gamma$ or $WWZ$
vertex, leads to an overall sign change.  Of course the same argument applies if 
the quarks are replaced by neutrinos and charged leptons.  An additional 
coupling factor is extracted for convenience, leaving finally the 
process-independent scalar form factors $\Delta\Gamma_L^{\gamma}(\qsq)$, 
$\Delta\Gamma_L^{Z}(\qsq)$ and $\Delta\Gamma_L^{W}(\qsq)$ on the 
right-hand side.  Finally, we form the combinations
\begin{mathletters}
\label{define-pibar}
\begin{eqnarray}
\label{define-piggbar}
\Delta\barpitgg(\qsq) & = & 
\Delta\pitgg(\qsq) - 2\hats\qsq\Delta\Gamma_L^\gamma(\qsq) 
\;,\\
\label{define-pigzbar}
\Delta\barpitgz (\qsq) & = & 
\Delta\pitgz(\qsq) - \hats\qsq\Delta\Gamma_L^Z(\qsq) - 
\hatc(\qsq-\mzsq)\Delta\Gamma_L^\gamma(\qsq)  
\;,\\
\label{define-pizzbar}
\Delta\barpitzz(\qsq) & = & 
\Delta\pitzz(\qsq) - 2\hatc(\qsq-\mzsq)\Delta\Gamma_L^Z(\qsq)
\;,\\
\label{define-piwwbar}
\Delta\barpitww(\qsq) & = &  
\Delta\pitww(\qsq) - 2(\qsq-\mwsq)\Delta\Gamma_L^W(\qsq)\;,
\end{eqnarray}
\end{mathletters}
where the $\overline{\Pi}_T^{V_1V_2}(\qsq)$'s on the left-hand side are now 
gauge-invariant expressions.

The contributions of these two-point-functions to four-fermion amplitudes is
generally summarized by a set of parameters such as the $S$, $T$ and $U$
parameters of Ref.~\cite{pt90andpt92} or an  
equivalent set\cite{triplets}.  Following Ref.~\cite{hhkm94} we define
\begin{mathletters}
\label{define-stu}
\begin{eqnarray}
\label{define-s}
\alpha\Delta S & = & 4\hatssq\hatcsq \bigg\{
  -\Delta\barpitzzz(0) 
  + \frac{\hatcsq-\hatssq}{\hats\hatc}\Delta\barpitggz(\mzsq) 
  + \Delta \barpitggg (\mzsq)\bigg\}
\;,\\
\label{define-t}
\alpha\Delta T & = & \frac{\Delta\barpitzz(0)}{\mzsq}
                    -\frac{\Delta\barpitww(0)}{\mwsq}
\;,\\
\label{define-u}
\alpha\Delta U & = & 4\hatssq \bigg\{
         \hatcsq\Delta\barpitzzz(0) 
        -\Delta\barpitwww(0) 
        +\hatssq\Delta\barpitggg(\mzsq)
        +2\hats\hatc\Delta\barpitggz(\mzsq) 
   \bigg\}
\;,
\end{eqnarray}
\end{mathletters}
where
\begin{equation}
\label{defn-pitv}
\overline{\Pi}^{V_1V_2}_{T,V_3}(\qsq) = {\overline{\Pi}^{V_1V_2}_T(\qsq) - 
\overline{\Pi}^{V_1V_2}_T(m^2_{V_3})\over \qsq -
m^2_{V_3}}\;.
\end{equation}
Notice the different subscripts on the left-hand and right-hand sides of 
Eqn.~(\ref{defn-pitv}).  

Several points concerning the usage of $\Delta S$, $\Delta T$ and $\Delta U$
should be made.  First of all, we may expand the $\overline{\Pi}$ functions
in a power series in $\qsq$ according to 
\begin{equation}
\label{expansion}
\Delta\overline{\Pi}^{V_1V_2}_T(\qsq) = A^{V_1V_2} + \qsq\,B^{V_1V_2} 
   + (\qsq)^2\,C^{V_1V_2} + ... \;.
\end{equation}
If we include only the $A$ and $B$ coefficients in our expansion, then, 
considering all four $\overline{\Pi}$ functions, there are a total of 
eight constant coefficients.  By a Ward identity, 
$A^{\gamma\gamma}= A^{\gamma Z} = 0$.  Using the three physical input parameters
(for which we choose $\alpha$, $m_Z$ and $G_F$) eliminates three more, 
leaving three parameters, {\em i.e.} $\Delta S$, $\Delta T$ and $\Delta U$.
In particular we expect that all nondecoupling effects are absorbed in these 
three parameters.  

Of course, as we go beyond the $A$ and $B$ coefficients in
Eqn.~(\ref{expansion}) we expect that $\Delta S$, $\Delta T$ and $\Delta U$
are insufficient to include all possible effects.  In particular, if in 
Eqn.~(\ref{expansion}) we include 
the $C$ terms, we expect an additional four parameters.  With each additional 
new term we expect four more parameters.  However, if we introduce four new
form factors that run with $\qsq$, then $S$, $T$ and $U$ plus these four are 
sufficient regardless of how many terms we retain in Eqn.~(\ref{expansion}).

For convenience in organizing our overall analysis we introduce four such 
running coefficients which may be expressed as linear combinations of the 
$\overline{\Pi}$ functions.  While these quantities are useful as a  means of 
organizing our calculations, we will later replace them with something else.
We write
\begin{mathletters}
\label{define-RVV}
\begin{eqnarray}
\label{define-Rgg}
\alpha\rgg(\qsq) & = & \frac{1}{\qsq}
\bigg[ \barpitggg(\qsq) - \barpitgg(0) \bigg]
\;,\\
\label{define-Rgz}
\alpha\rgz(\qsq) & = & \frac{1}{\qsq-\mzsq} 
\bigg[ \barpitggz(\qsq) - \barpitggz(\mzsq) \bigg]
\;,\\
\label{define-Rzz}
\alpha\rzz(\qsq) & = & \frac{1}{\qsq} 
\bigg[\barpitzzz(\qsq) - \barpitzzz(0) \bigg]
\;,\\
\label{define-Rww}
\alpha\rww(\qsq) & = & \frac{1}{\qsq} 
\bigg[ \barpitwww(\qsq) - \barpitwww(0) \bigg]
\;.
\end{eqnarray}
\end{mathletters}
These quantities are generated directly by energy-dimension-six operators 
or loop effects.
In Ref.~\cite{beyond} three parameters, $V$, $W$ and $X$, were introduced.  
They are equivalent to $\rzz(\mzsq)$, $\rww(\mwsq)$ and $\rgz(0)$.  
Because current experiments are not sensitive to the fourth parameter,
the authors of that work did not introduce a parameter equivalent to $\rgg$. 

Expressed in terms of the seven parameters $\Delta S$, $\Delta T$, 
$\Delta U$, $\Delta\rgg$, $\Delta\rgz$, $\Delta\rzz$ 
and $\Delta\rww$, we introduce four effective charges\cite{hhkm94} via
\begin{mathletters}
\label{barred-charges}
\begin{eqnarray}
\label{alpha-bar}
\Delta\baralphaqsq & = & 
-\hatalpha^2 \qsq \Delta\rgg(\qsq)
\;,\\
\label{gz-bar}
\Delta\bargzsqqsq & = & 
\hatalpha\hatgzsq \bigg[\Delta T - \qsq\Delta\rzz(\qsq)\bigg]
\;,\\
\label{s2-bar}
\Delta\barssqqsq & = & 
\frac{\hatssq\hatcsq}{\hatcsq-\hatssq}\Bigg[
\frac{\Delta\baralpha(\mzsq)}{\hatalpha}
-\frac{\bargzsq(0)}{\hatgzsq}
+ \frac{\hatalpha\Delta S}{4\hatssq\hatcsq}\Bigg]
+ \hatalpha\hats\hatc\,\Big(\qsq-\mzsq\Big)\Delta\rgz(\qsq)
\;,\\
\label{gw-bar}
\Delta\bargwsqqsq & = & 
- \hatgsq\frac{\Delta\barssq(\mzsq)}{\hatssq}
+ \hatgsq\frac{\Delta\baralpha(\mzsq)}{\hatalpha}
+ \hatalpha\hatgsq\frac{\Delta S + \Delta U}{4\hatssq}
- \hatalpha\hatgsq\qsq\Delta\rww(\qsq)
\;.
\end{eqnarray}
\end{mathletters}
When going beyond effects which may be summarized by $\Delta S$, 
$\Delta T$ and $\Delta U$, we find that it is most pragmatic to simply use 
the above effective charges.  This avoids a proliferation of new parameters,
a subset of which must be allowed to run anyway.  Furthermore, the physical 
interpretation of the effective charges is straightforward\cite{prw96}.
Notice that Eqns.~(\ref{alpha-bar})-(\ref{gw-bar}) must be calculated 
sequentially as presented.

Finally, we must consider process dependent vertex and box corrections.
In general there could be a large number of such corrections.  However, for 
the current analysis, the only non-SM vertex correction with which we must 
be concerned is the correction to the $Zb\barb$ vertex arising from
the graph of Fig.~\ref{fig-vertex}(b) with an internal top-quark line.  We
introduce a form factor\cite{hhkm94}, $\bardeltab(\qsq)$, which changes 
the SM Feynman rule for the $Zb\barb$ vertex to 
\begin{equation}
\label{delta-zbb}
-i \hatgz\gamma^\mu\bigg\{ -\hatssq Q_b P_+ + 
\Big[ \big(1 + \bardeltab(\qsq)\big) I_3 - \hatssq Q_b\Big]P_-\bigg\},
\end{equation}
where the projection operators are defined by $P_\pm = (1\pm\gamma_5)/2$,
and $Q_b = -1/3$ and $I_3 = -1/2$ are the charge and weak-isospin quantum 
numbers of the b quark.  Using the decomposition $\bardeltab = 
\overline{\delta}_{b\;\rm SM} + \Delta \bardeltab$, the first term 
contains the entire SM vertex correction (minus the pinch term) that
multiplies $I_3$, and the `$\Delta$' term is the contribution of  
Fig.~\ref{fig-vertex}(b) (also minus the pinch term).

In the next two sections we discuss possible parameterizations of
new physics effects and apply the  formalism
developed above to these scenarios.

\section{The light-Higgs scenario}\label{sec-linear}

Assuming the existence of a physical Higgs boson new physics may be described by 
an SU(2)$\times$U(1) gauge-invariant effective Lagrangian of the form
\begin{equation}
\label{leff-lr}
\call_{\rm eff}^{\rm linear} = \call_{\rm SM} + \sum \frac{f_i}{\Lamsq}
{\cal O}_i + \cdots \;.
\end{equation}
The first term is the usual SM Lagrangian which includes a complete set of 
gauge-invariant \oefour operators and explicitly includes operators 
involving the SM Higgs doublet, $\Phi$.  The second term 
constitutes a complete set of \oesix operators; each \oesix
operator, ${\cal O}_i$, is multiplied by a dimensionless numerical coefficient, 
$f_i$,  and is explicitly suppressed by inverse powers of the scale of new 
physics, $\Lambda$, such that the overall energy dimension equals four. In 
general a very large number of new operators could 
contribute\cite{bw86,bs83andllr86}.  
However, including only those purely bosonic operators which conserve CP, only 
twelve C- and P-conserving operators remain\cite{hisz}.  The explicit 
expressions for these operators are relegated to Appendix~\ref{app-linear}.

Four operators $\odw$, $\odb$, $\obw$ and $\opone$  (with associated 
coefficients 
$\fdw$, $\fdb$, $\fbw$ and $\fpone$ respectively) are especially important for 
their contributions at the tree level to the two-point-functions of the 
electroweak gauge bosons\cite{hisz,gw91,hms95}, although $\odw$ and $\obw$ 
contribute to nonstandard $WW\gamma$ and $WWZ$ couplings as well.  Three 
operators, $\ow$, $\ob$ and $\owww$ (with associated coefficients $\fw$, $\fb$, 
$\fwww$) are significant because they contribute at the tree level to 
nonstandard $WW\gamma$ and $WWZ$ interactions without an associated tree-level 
contribution to the two-point functions. While the tree-level contributions to 
the gauge-boson two-point-functions of the two operators $\oww$ and $\obb$ 
(with respective coefficients $\fww$ and $\fbb$) may be removed by a trivial
redefinition of fields and couplings\cite{hisz,hhis96}, these operators are 
still interesting for their contributions to $H\gamma\gamma$ and $HZ\gamma$ 
vertices\cite{hsz93}.  The operator $\opfour$ makes a contribution to the $ZZ$ 
and $WW$ two-point-functions, but the contributions cancel in physical 
observables.  Hence, $\optwo$, $\opthree$ and $\opfour$ contribute only to 
Higgs-boson self-interactions and are of no further interest in the current 
context.  Additional details may be found in Refs.~\cite{hisz,hms95,hhis96}.

We will use the effective charges calculated to the leading order in each 
operator.  In other words, only the tree-level contributions of $\odw$, $\odb$, 
$\obw$ and $\opone$ will be included while $\ow$, $\ob$, $\owww$, $\oww$ and
$\obb$ contribute through loop diagrams.  All calculations in 
this section were performed in $R_\xi$ gauge.  We calculate the loop graphs in 
$d=4-2\epsilon$ dimensions and identify the poles at $d=4$ with logarithmic 
divergences and make the identification
\begin{equation}
\label{log-divergence}
\frac{1}{\epsilon}(4\pi)^\epsilon\Gamma(1+\epsilon)
\rightarrow \loglammu
\quad ,
\end{equation} 
where $\mu$ is an arbitrary renormalization scale.  We have retained only the 
logarithmic terms and terms which grow with the mass of the Higgs boson, $m_H$.  
Combining the results of Refs.~\cite{hisz,hms95}  we may write the 
solution as
\begin{mathletters}
\label{results-lin}
\begin{eqnarray}
\nonumber
\alpha\Delta S & = & -\hatesq\frac{v^2}{\Lamsq}\fbw 
  -\frac{1}{6}\frac{\hatesq}{16\pi^2}\Bigg\{
	3(\fw+\fb)\frac{\mhsq}{\Lamsq}\bigg[\loglammh + \frac{1}{2}\bigg] 
\\ \nonumber && \mbox{}
  +2 \Big[(5\hatcsq-2)\fw-(5\hatcsq-3)\fb\Big]\frac{\mzsq}{\Lamsq}\loglammh
\\ \nonumber && \mbox{}
  - \Big[(22\hatcsq-1)\fw-(30\hatcsq+1)\fb\Big]\frac{\mzsq}{\Lamsq}\loglammz
\\ && \mbox{}
  - 24 (\hatcsq\fww+\hatssq\fbb)\frac{\mzsq}{\Lamsq}\loglammh
  + 36\hatgsq\fwww\frac{\mwsq}{\Lamsq}\loglammz \Bigg\}
\label{deltas-lin}
\;, \\
\nonumber
\alpha\Delta T & = & -\frac{1}{2}\frac{v^2}{\Lamsq}\fpone 
   -\frac{3}{4\hatcsq}\frac{\hatesq}{16\pi^2}\Bigg\{ 
   \fb\frac{\mhsq}{\Lamsq}\bigg[\loglammh + \frac{1}{2}\bigg] 
\\ \nonumber && \mbox{}
   + (\hatcsq\fw+\fb)\frac{\mzsq}{\Lamsq}\loglammh
\\ && \mbox{}
   + \Big[2\hatcsq\fw + (3\hatcsq-1)\fb\Big]\frac{\mzsq}{\Lamsq}\loglammz
   \Bigg\}
\label{deltat-lin}   
\;, \\
\nonumber
\alpha\Delta U & = & 8\hatesq\hatssq\frac{\mzsq}{\Lamsq}\fdw  +
\frac{1}{3}\frac{\hatesq\hatssq}{16\pi^2}\Bigg\{
(-4\fw+5\fb)\frac{\mzsq}{\Lamsq}\loglammh
\\ && \mbox{}
+(2\fw-5\fb)\frac{\mzsq}{\Lamsq}\loglammz\Bigg\}
\label{deltau-lin}\;, \\
\label{deltargg-lin}
\alpha\Delta\rgg(\qsq) & = & \frac{2\hatesq}{\Lamsq}(\fdw+\fdb)
- \frac{1}{6\Lamsq}\frac{\hatesq}{16\pi^2}(\fw+\fb)\loglammz
\;, \\
\nonumber
\alpha\Delta\rgz(\qsq) & = & 
\frac{2\hate\hatgz}{\Lamsq}(\hatcsq\fdw-\hatssq\fdb)+ 
\frac{1}{24\Lamsq}\frac{\hate\hatgz}{16\pi^2}
\Bigg\{(\fb-\fw)\loglammh
\\\label{deltargz-lin} && \mbox{} -
\Big[(4\hatcsq-1)\fw+(4\hatcsq-3)\fb\Big]\loglammz\Bigg\} \;, \\
\nonumber
\alpha\Delta\rzz(\qsq) & = & \frac{2\hatgzsq}{\Lamsq}(\hatc^4\fdw+\hats^4\fdb)
- \frac{1}{12\Lamsq}\frac{\hatgzsq}{16\pi^2}
\Bigg\{(\hatcsq\fw+\hatssq\fb)\loglammh
\\&&\mbox{} +(\hatcsq-\hatssq)(\hatcsq\fw-\hatssq\fb)\loglammz\Bigg\} 
\label{deltarzz-lin} \;, \\
\label{deltarww-lin}
\alpha\Delta\rww(\qsq) & = & \frac{2\hatgsq}{\Lamsq}\fdw 
- \frac{1}{12\Lamsq}\frac{\hatgsq}{16\pi^2}\fw
 \Bigg\{\loglammh+\loglammz \Bigg\} \;,
\end{eqnarray}
\end{mathletters}
where $v = 246$~GeV is the vacuum expectation value of the Higgs field.  From 
these expressions we may immediately calculate the effective charges of 
Eqns.~(\ref{barred-charges}).  
Everywhere we have made the assignment $\mu^2 = \mzsq$.  

Finally, we calculate the $Zb\barb$ vertex form factor,
\begin{equation}
\label{delta-zbb-linear}
\Delta\bardeltab(\qsq) = -\frac{\hatalpha}{16\pi\hatssq}
\frac{m_t^2}{\Lamsq}\Bigg\{\frac{\qsq}{\mwsq}\frac{(\hatcsq\fw-\hatssq\fb)}{2}
+3\fw\Bigg\}\loglammz\;.
\end{equation}
This result agrees with Ref.~\cite{dv95}, as discussed below.  Such effects 
have also been considered in Ref.~\cite{bgklm94}.  Recall that we began 
with operators composed only of bosonic fields.  A nonzero value for 
$\Delta\bardeltab$ indicates that mixed bosonic-fermionic operators have 
been radiatively generated.  

\section{The electroweak chiral Lagrangian}\label{sec-nonlinear}

Next we address the nonlinear realization of the symmetry breaking sector.
In the notation of Ref.~\cite{lon80,lon81,aw93} we present the chiral 
Lagrangian,
\begin{equation}
\label{leff-nlr}
\call_{\rm eff}^{\rm nlr} = \call_{\rm SM}^{\rm nlr} + 
\sum \call_{i} + \cdots \;.
\end{equation}
We use the superscript `nlr', denoting `nonlinear realization'.  Again the 
first term is the SM Lagrangian, but in this case no physical Higgs boson
is included.  Hence $\call_{\rm SM}^{\rm nlr}$  is nonrenormalizable.
The first non-SM terms are energy \oetwo and \oefour
operators which are not manifestly suppressed by explicit powers of some high 
scale.  There are twelve such operators which conserve CP; eleven of these 
separately conserve C and P.  For explicit notation see 
Appendix~\ref{app-nonlinear}.

Three of the operators, $\call_{1}^\prime$, $\call_{1}$ and 
$\call_{8}$, contribute at the tree-level to the gauge-boson 
two-point-functions; $\call_{1}$ and $\call_{8}$ also contribute to 
nonstandard $WW\gamma$ and $WWZ$ couplings.  Three operators, $\call_{2}$, 
$\call_{3}$ and $\call_{9}$, contribute to $WW\gamma$ and $WWZ$ couplings 
without making a tree-level contribution to the gauge-boson propagators.
Unlike the light-Higgs scenario, several operators, $\call_{4}$-$\call_{7}$
and $\call_{10}$, contribute only to quartic vertices.  Several operators 
violate the custodial symmetry, SU(2)$_{\rm C}$.  They are $\call_{1}^\prime$,
$\call_{6}$, $\call_{7}$, $\call_8$, $\call_{9}$ and $\call_{10}$.
$\call_1^\prime$ is \oetwo in the energy expansion and violates the 
custodial $SU(2)_{\rm C}$ symmetry even in the absence of gauge couplings.  
Finally, 
$\call_{11}$ is special in the sense that it conserves CP while it violates P.
This operator contributes to the four-fermion matrix elements through a myriad of
process-dependent vertex corrections.  For this reason it is not easily included 
in the current analysis.  Its contributions to low-energy and $Z$-pole data were 
discussed in Ref.~\cite{dv94}.

Each operator in Eqns.~(\ref{twelve-chiral}) has a counterpart in the linear 
realization of SSB\cite{hhis96,hag95}.  Four of these counterparts are 
\oesix operators and appear in Eqns.~(\ref{twelve}).  We 
make the correspondence,
\begin{mathletters}
\label{relate4}
\begin{eqnarray}
\label{l1p0p1}
\call_1^\prime & = & - \frac{4 \beta_1}{v^2} {\cal O}_{\Phi,1} \;, \\
\label{l1obw}
\call_1 & = &  \frac{4 \alpha_1}{v^2} {\cal O}_{BW} \;, \\
\label{l2ob}
\call_2 & = &  \frac{8 \alpha_2}{v^2} {\cal O}_{B} \;, \\
\label{l3ow}
\call_3 & = &   \frac{8 \alpha_3}{v^2} {\cal O}_{W} \;.
\end{eqnarray}
\end{mathletters}

The two-point-functions in the context of the chiral Lagrangian were calculated 
in the unitary gauge by the authors of Ref.~\cite{dv95}.  Some contributions were
also checked by applying Eqns.~(\ref{relate4}) to the results of 
Ref.~\cite{hisz} and carefully removing all Higgs boson contributions.  The 
contributions of those operators which contribute only to the quartic vertices 
were also obtained in Ref.~\cite{beg96}.\footnote{The purely quartic operators 
contribute only to the $T$ parameter via Eqn.~(\protect\ref{deltat-nl}).  Our 
results disagree with those of Ref.~\protect\cite{beg96} for the contributions
of $\call_4$, $\call_5$ and $\call_7$, while we have differing conventions 
for $\call_{10}$.}
We summarize our one-loop results as
\begin{mathletters}
\label{results-nl}
\begin{eqnarray}
\nonumber
\alpha\Delta S & = & 
\frac{\hatalpha}{12\pi}\loglamhatmu
- 4\hatesq \alpha_1
\\\label{deltas-nl} &&\mbox{}
-\frac{\hatgsq\hatesq}{16\pi^2}
       \bigg[\frac{1+30 \hatcsq}{3\hatcsq} \alpha_2 
       +\frac{1-22\hatcsq}{3\hatcsq} \alpha_3
       +\frac{1+6\hatcsq}{3\hatcsq} \alpha_9
        \bigg]\loglammz
\;,\\ \nonumber
\alpha\Delta T & = & 
\frac{3\hatalpha}{16\pi\hatcsq}
\loglamhatmu + 2\beta_1 
\\\nonumber &&\mbox{}
- \frac{\hatgsq\hatgzsq}{16\pi^2}
\bigg[ \frac{3\hatssq(3\hatcsq-1)}{2\hatcsq}\alpha_2 
      + 3\hatssq \alpha_3 
      + \frac{15\hatssq(\hatcsq+1)}{4\hatcsq}\alpha_4
      + \frac{3\hatssq(\hatcsq+1)}{2\hatcsq}\alpha_5
\\ \label{deltat-nl} && \mbox{}
      + \frac{3(2\hatc^4+11)}{4\hatcsq}\alpha_6
      + \frac{6(\hatc^4+1)}{\hatcsq}\alpha_7
      + \frac{3\hatssq}{2}\alpha_9
      + \frac{9}{\hatcsq}\alpha_{10}
       \bigg]\loglammz
\;,\\ \nonumber
\alpha\Delta U & = & -4\hatesq \alpha_8
        + \frac{\hatg^4}{16\pi^2} \frac{2\hatssq}{3\hatcsq}\bigg[
        -\hatssq(2\hatcsq+3)\alpha_2 
\\ \label{deltau-nl} && \mbox{}
        + 2\hatssq (2 \hatssq + \hatcsq)\alpha_3
        + (2 \hatc^4 - 15 \hatcsq + 1) \alpha_9 \bigg]
        \loglammz
\;,\\
\label{deltargg-nl}
\alpha\Delta\rgg(\qsq) & = & 
   - \frac{\hatesq\hatgsq}{16\pi^2}\frac{1}{3\mwsq}
   \Big(\alpha_2+\alpha_3+\alpha_9 \Big)\loglammz \;,\\
\label{deltargz-nl}
\alpha\Delta\rgz(\qsq) & = & 
    -\frac{\hate\hatgz\hatgsq}{16\pi^2}\frac{4\hatcsq-1}{12\mwsq}
   \Big(\alpha_2+\alpha_3+\alpha_9 \Big)\loglammz \;,\\
\label{deltarzz-nl}
\alpha\Delta\rzz(\qsq) & = & 
    -\frac{\hatgzsq\hatgsq}{16\pi^2}\frac{\hatcsq-\hatssq}{6\mwsq}
   \Big(-\hatssq\alpha_2+\hatcsq\alpha_3+\hatcsq\alpha_9 \Big)
        \loglammz \;,\\
\label{deltarww-nl}
\alpha\Delta\rww(\qsq) & = & 
    -\frac{\hatg^4}{16\pi^2}\frac{1}{6\mwsq}\alpha_3
        \loglammz
\;.
\end{eqnarray}
\end{mathletters}
As before, we have computed only the divergent contributions and replaced 
${1\over\epsilon}(4\pi)^\epsilon\Gamma(1+\epsilon) \rightarrow 
\log(\Lamsq/\mu^2)$ and have dropped all nonlogarithmic terms.\footnote{The 
contributions of the SU(2)$_{\rm C}$ conserving terms  can be obtained from 
the Appendix of Ref. \cite{dv95} by making the substitutions 
$L_{10}v^2/\Lamsq\rightarrow  \alpha_1$, 
$L_{9R}v^2/\Lamsq\rightarrow 2 \alpha_2$,
$L_{9L}v^2/\Lamsq\rightarrow 2 \alpha_3$,
$L_{2}v^2/\Lamsq\rightarrow  \alpha_4$,
$L_{1}v^2/\Lamsq\rightarrow  \alpha_5$.}
Furthermore we have chosen $\mu^2=\mzsq$.  Even
when all the $\alpha_i$ are zero, the expressions for $\Delta S$ and
$\Delta T$ are nonzero.  This is because the nonlinear Lagrangian contains 
singularities which in the SM would be cancelled by the contributions of the 
Higgs boson\cite{dob91}.  In these terms the renormalization scale, ${\hat\mu}$, 
is appropriately taken to be the same Higgs-boson mass we use to evaluate the SM 
contributions.

The next step is to use Eqn.~(\ref{barred-charges}) to calculate the effective 
charges.  However, the expressions become rather complicated, so we will leave 
them in the above form.  The nonzero expressions on the right-hand 
sides of Eqns.~(\ref{deltargg-nl})-(\ref{deltarww-nl}) are a clear indication 
that \oesix operators have been radiatively generated.  

To complete this section we present the calculation of $\Delta\bardeltab$ in the 
nonlinear model\cite{dv95}:
\begin{equation}
\label{delta-zbb-nl}
\Delta\bardeltab(\qsq) = 
-\frac{\hatalpha}{16\pi\hatssq}\frac{m_t^2}{\mzsq}
\hatgzsq\Bigg\{(-\hatssq\alpha_2+\hatcsq\alpha_3+\hatcsq\alpha_9)
 \frac{\qsq}{\mwsq} 
+6\alpha_3\Bigg\}\loglammz\;.
\end{equation}

\section{Non-Abelian gauge-boson vertices}\label{sec-nabgbv}

Much of the literature describes nonstandard $WW\gamma$ and $WWZ$ vertices 
via the phenomenological effective Lagrangian\cite{hpzh87}
\begin{eqnarray}
\nonumber
\lefteqn{\call_{WWV} =}\\&&  - i g_{WWV} \Bigg\{ 
g_1^V \Big( W^+_{\mu\nu} W^{- \, \mu} V^{\nu} 
  - W^+_{\mu} V_{\nu} W^{- \, \mu\nu} \Big) 
+ \kappa_V W_\mu^+ W_\nu^- V^{\mu\nu}
+ \frac{\lambda_V}{\mwsq} W^+_{\mu\nu} W^{- \, \nu\rho} V_\rho^{\; \mu}
 \Bigg\}
\;,\label{lagr-phenom}
\end{eqnarray}
where $V=Z,\gamma$, the overall coupling constants are 
$g_{WW\gamma} = \hate$ and 
$g_{WWZ} = \hatgz\hatcsq$.  The field-strength tensors include only
the Abelian parts, {\em i.e.}\/ 
$W^{\mu\nu} = \partial^\mu W^\nu - \partial^\nu W^\mu$
and $V^{\mu\nu} = \partial^\mu V^\nu - \partial^\nu V^\mu$.  
In Eqn.~(\ref{lagr-phenom}) we have retained only the terms which separately 
conserve C and P (since that is all that we retain in the previous sections).

In the light-Higgs scenario, if we neglect those operators which contribute
to gauge-boson two-point-functions at the tree level, we may write \cite{hisz93}:
\begin{mathletters}
\label{hisz-relations}
\begin{eqnarray}
\label{g1z-hisz}
\gonez(\qsq) & = & 1 + \frac{1}{2}\frac{\mzsq}{\Lamsq}\fw \;, 
\\
\label{kappagamma-hisz}
\kapgam(\qsq) & = & 1 + \frac{1}{2}\frac{\mwsq}{\Lamsq}\Big(\fw + \fb\Big) \;, 
\\
\label{kappaz-hisz}
\kapz(\qsq) & = & 1 + \frac{1}{2}\frac{\mzsq}{\Lamsq}
  \Big(\hatcsq\fw - \hatssq\fb\Big)\;, 
\\
\label{lambda-hisz}
\lamgam(\qsq) = \lamz(\qsq) & = & \frac{3}{2}\hatgsq\frac{\mwsq}{\Lamsq}\fwww \;.
\end{eqnarray}
\end{mathletters}
Hence, truncating the gauge-invariant expansion of Eqn.~(\ref{leff-lr}) at the 
level of \oesix operators produces nontrivial relationships 
between the nonstandard couplings.  These relationships are broken by the 
inclusion of \oeeight operators\cite{hisz93}.

We present similar equations arising from the electroweak chiral 
Lagrangian to \oefour in the energy expansion\cite{hhis96,dv94,fer93}:
\begin{mathletters}
\label{nl-hisz-relations}
\begin{eqnarray}
\label{nl-g1z-hisz}
\gonez(\qsq) & = & 1 + \hatgzsq\alpha_3 \;,
\\
\label{nl-kappagamma-hisz}
\kapgam(\qsq) & = & 1 + \hatgsq \Big( \alpha_2 + \alpha_3 + \alpha_9 \Big)\;, 
\\
\label{nl-kappaz-hisz}
\kapz(\qsq) & = & 1 + \hatgzsq\Big( 
-\hatssq\alpha_2  + \hatcsq\alpha_3 + \hatcsq\alpha_9 \Big)  \;, \\
\label{nl-lambda-hisz}
\lamgam(\qsq) & =& \lamz(\qsq)   \approx  0 \;.
\end{eqnarray}
\end{mathletters}
If we impose the custodial SU(2)$_{\rm C}$ symmetry on the new physics, then we 
may neglect the $\alpha_9$ terms.  In this case the correlations which 
exist in the light-Higgs scenario exist here as well.  Again, these relations
are violated by higher-order effects.  Eqn.~(\ref{nl-lambda-hisz}) reflects
our prejudice that the $\lambda_V$ couplings, being generated by 
\oesix operators while the other couplings are generated by 
\oefour operators, should be relatively small. 

Current data is sensitive to gauge-boson propagator effects, but measurements of 
$WW\gamma$ and $WWZ$ couplings are rather crude.  Until the quality of the 
latter measurements approaches the quality of the former, the approximations of 
this section are valid. 

\section{Numerical Analysis and Discussion}\label{sec-numerical}

We begin this section by summarizing the results of a recent global 
analysis\cite{hhm96}.  For measurements on the $Z$-pole,
\begin{equation}
\label{z-pole-data}
\left. \begin{array}{l}
\overline{g}_Z^2(\mzsq) = 0.55557 - 0.00042
\frac{\alpha_s + 1.54 \overline{\delta}_b(\mzsq) - 0.1065}{0.0038} \pm 0.00061 
\\
\overline{s}^2(\mzsq) = 0.23065 + 0.00003
\frac{\alpha_s + 1.54 \overline{\delta}_b(\mzsq) - 0.1065}{0.0038} \pm 0.00024
\end{array}\right\}\makebox[.2cm]{}
\rho_{\rm corr} = 0.24 \;.
\end{equation}
The correlation between the two measurements is given by 
$\rho_{\rm corr}$\cite{pdg96}.
Recall $\overline{\delta}_b(\mzsq) = \overline{\delta}_{b\; \rm SM}(\mzsq)
+ \Delta\overline{\delta}_b(\mzsq)$.  Combining the $W$-boson mass 
measurement ($m_W = 80.356\pm 0.125$~GeV) with the input parameter $G_F$,
\begin{eqnarray}
\label{w-data}
\overline{g}_W^2(0) = 0.4237 \pm 0.0013\;.
\end{eqnarray}
And finally, from the low-energy data,
\begin{equation}
\label{le-data}
\left. \begin{array}{l}
\overline{g}_Z^2(0) = 0.5441 \pm 0.0029 \\
\overline{s}^2(0) = 0.2362 \pm 0.0044
\end{array}\right\}\makebox[0.2cm]{} \rho_{\rm corr} = 0.70 \;.
\end{equation}
We combine these results with the analytical results of the previous sections  
 to perform a $\chi^2$ analysis and obtain limits on the 
coefficients of both the linear and nonlinear models.

\subsection{Results for the linear model}

For those operators that contribute at the tree level the bounds which we obtain 
are straightforward and unambiguous.  For these operators we present the 
 fits along with the complete one-sigma errors\cite{hhis96}
\begin{equation}
\label{fit-now}
\left.\begin{array}{lll}
\fdw & = & -0.32 + 0.0088 \,x_H - 0.55 \,x_t \pm 0.44 \;,\\
\fdb & = & -14 \pm 10 \;,\\
\fbw & = & 3.7  + 0.085 \,x_H \pm 2.4 \;,\\
\fpone & = & 0.30 - 0.028 \,x_H + 0.32 \,x_t  \pm 0.16 \;,
\end{array}\right.
\end{equation}
and the full correlation matrix
\begin{equation}
\rho_{\rm corr} = 
\left(\makebox[-0.3cm]{}
\begin{array}{lddd}
\dec 1. & \dec $-$0.191 & \dec    0.055 & \dec $-$0.237  \\
        & \dec    1.    & \dec $-$0.988 & \dec $-$0.884  \\
        &               & \dec    1.    & \dec    0.943  \\
        &               &               & \dec    1.
\end{array} \right) 
\end{equation}
where 
\begin{equation}
x_t = \frac{m_t-175{\rm\; GeV}}{100{\rm\; GeV}}\;,  \makebox[1cm]{}
x_H = \ln \frac{m_H}{100{\rm\; GeV}} \;,
\end{equation}
and $\Lambda = 1$~TeV.  The parameterization of the central values is good to a 
few percent of the one-sigma errors in the range 
$150{\rm\; GeV}<m_t<190{\rm\; GeV}$ and $60{\rm\; GeV}<m_H<800{\rm\; GeV}$; for 
these four parameters the dependencies upon $m_H$ and $m_t$ 
arise from SM contributions only.  These bounds will improve with the analysis 
of LEP~II data; the process $e^+e^-\rightarrow W^+W^-$ is sensitive to $\fbw$ 
even at the level of the current constraints\cite{hhis96}, and all of the bounds 
improve significantly when LEP~II data for two-fermion final states are combined
with the current analysis\cite{hms95}.

The constraints on the remaining parameters are more subject to interpretation.
We make a distinction between those operators which first contribute to 
four-fermion amplitudes at the tree level and those which first contribute at 
the loop level.  Without an explicit model from which to calculate, it is most 
natural to assume that all of the coefficients are generated with similar 
magnitudes\cite{drghm92}.  Generally the contributions which first arise at one 
loop are suppressed by a factor  of $1/16\pi^2$ relative to tree-level effects; 
hence the contributions of operators first contributing at the loop level tend 
to be obscured.  Furthermore, outside of a particular model it is impossible to 
predict the interference between tree-level and loop-level diagrams as well as
possible cancellations among the various loop-level contributions.  For the time 
being we will proceed by considering the effects of only one operator at a time.
The results are presented in Table~\ref{table-linear-singly}.
\begin{table}[tbh]
\begin{tabular}{||l||c|c|c|c||}
      & $m_H = $~75~GeV & $m_H = $~200~GeV & $m_H = $~400~GeV & $m_H = $~800~GeV 
\\ \hline\hline
$\fwww$ & -21$\pm$10 & 5$\pm$10 & 24$\pm$10 & 43$\pm$10
\\ \hline
$\fw$   & 2.4$\pm$3.2 & -5.0$\pm$3.8 & -7.5$\pm$4.5 & 2.2$\pm$3.8 \\ \hline
$\fb$   & -5.0$\pm$9.8 & 7.1$\pm$7.5 & 0.78$\pm$4.2 & -3.0$\pm$2.8 \\ \hline
$\fww$  & 12.5$\pm$6.0 & -4.8$\pm$9.7 & -39$\pm$17 & -289$\pm$70 \\ \hline
$\fbb$  & 42$\pm$20 & -16$\pm$32 & -131$\pm$57 & -960$\pm$233 \\
\end{tabular}
\caption{One-sigma fits for the coefficients of $\owww$, $\ow$, $\ob$, 
$\oww$ and $\obb$ for $\Lambda = 1$~TeV.  In the analysis only one coupling
at a time is allowed to deviate from zero.}
\label{table-linear-singly}
\end{table}
In general we find consistency with the SM for a relatively light, 
100~GeV-200~GeV Higgs boson.  For $\fwww$ the central values depend upon 
$m_H$ only through SM contributions, and the one-sigma error is independent of 
$m_H$.  For $\fw$, $\fb$, $\fww$ and $\fbb$ the dependence on the Higgs-boson 
mass is from both SM and non-SM contributions, and both the central values and 
the errors are complicated functions of $m_H$.

\begin{figure}[htb]
\begin{center}
\leavevmode\psfig{file=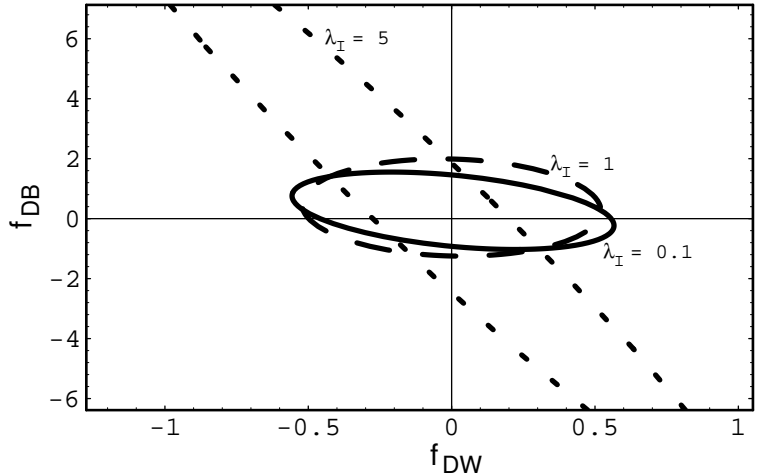,angle=0,width=10cm,silent=0}
\end{center}
\caption{Constraints at the 95\% confidence level in the $\fdw$--$\fdb$ plane 
with $\fww = 20\lambda_I\fdw$ and $\fbb = 20\lambda_I\fdb$ for $\Lambda = 1$~TeV,
$m_t = 175$~GeV and $m_H = 200$~GeV.  The solid, dashed and dotted curves 
correspond to $\lambda_I = $~0.1, 1 and 5 respectively.}
\label{fig-fdwfdb} 
\end{figure}

It is also possible that there is a hierarchy among the coefficients,
some being relatively large while others are relatively suppressed.  In the 
current discussion it is especially interesting if all of the operators 
with non-negligible coefficients contribute only at the loop level.  Indeed such 
a scenario is possible.  Consider, for example, the simple model described by 
the Lagrangian\cite{dz}
\begin{equation}
{\cal L} = {\cal L}_{\rm SM} + \big(D_\mu\phi\big)^\dagger \big(D^\mu\phi\big)
- m_0^2\phi^\dagger\phi 
+ \lambda_I \big(\phi^\dagger\phi\big) \big(\Phi^\dagger\Phi\big) 
- \lambda^\prime\big(\phi^\dagger\phi\big)^2,
\end{equation}
where $\Phi$ is the SM Higgs doublet, $\phi$ is a new heavy scalar with isospin
$I$ and hypercharge $Y$.  The self-coupling of the new scalar is given by 
$\lambda^\prime$, and $\lambda_I$ denotes the interaction strength.  The physical
mass of the heavy scalar is given by $m^2 = m_0^2 - \lambda v^2/2$.  The above 
Lagrangian generates the following nonzero couplings:
\begin{mathletters}
\label{new-scalar}
\begin{eqnarray}
\label{new-scalar-fdw}
\frac{\fdw}{\Lambda^2} & = & \frac{1}{16\pi^2}\frac{1}{m^2}
\frac{I(I+1)(2I+1)}{180} \;,\\
\label{new-scalar-fdb}
\frac{\fdb}{\Lambda^2} & = & \frac{1}{16\pi^2}\frac{1}{m^2}
\frac{Y^2(2I+1)}{60} \;,\\
\label{new-scalar-fww}
\frac{\fww}{\Lambda^2} & = & \frac{1}{16\pi^2}\frac{1}{m^2}\lambda_I
\frac{I(I+1)(2I+1)}{9} = 20 \lambda_I \frac{\fdw}{\Lambda^2} \;,\\
\label{new-scalar-fbb}
\frac{\fbb}{\Lambda^2} & = & \frac{1}{16\pi^2}\frac{1}{m^2}\lambda_I
\frac{Y^2(2I+1)}{3} = 20 \lambda_I \frac{\fdb}{\Lambda^2} \;.
\end{eqnarray}
\end{mathletters}
The remaining couplings remain explicitly zero.  It is immediately apparent
that, for large values of $\lambda_I$, the couplings $\fww$ and $\fbb$ may be 
large relative to $\fdw$ and $\fdb$.  (Of course for large $\lambda_I$ there 
may also be large corrections to the above relations.)  Unfortunately this 
scenario is numerically problematic.  If we are interested in the large coupling 
limit where $\fdw,\fdb \ll \fww,\fbb$ it is impossible to obtain any constraint 
at all.  This may be seen from Eqns.~\ref{results-lin}; the operators 
$\oww$ and $\obb$ contribute only through the $(\hatcsq\fww+\hatssq\fbb)$ term 
in $\Delta S$ of Eqn.~\ref{deltas-lin}.  (Notice also that $\fwww$ enters only
through $\Delta S$.)  In Fig.~\ref{fig-fdwfdb}
the solid, dashed and dotted curves represent $\lambda_I = $~0.1, 1 and 5 
respectively.  For the weak coupling ($\lambda_I = $~0.1) the contributions 
of $\fww$ and $\fbb$ are completely negligible.  For $\lambda_I = $~1 the 
effects of $\fww$ and $\fbb$ are competitive with those of $\fdw$ and $\fdb$.
Finally, when $\lambda_I = $~5 the fit is dominated by the strong anti-correlation
of $\fww$ and $\fbb$.  In the strong-coupling limit the very eccentric ellipse
approaches a line.

\begin{figure}[htb]
\begin{center}
\leavevmode\psfig{file=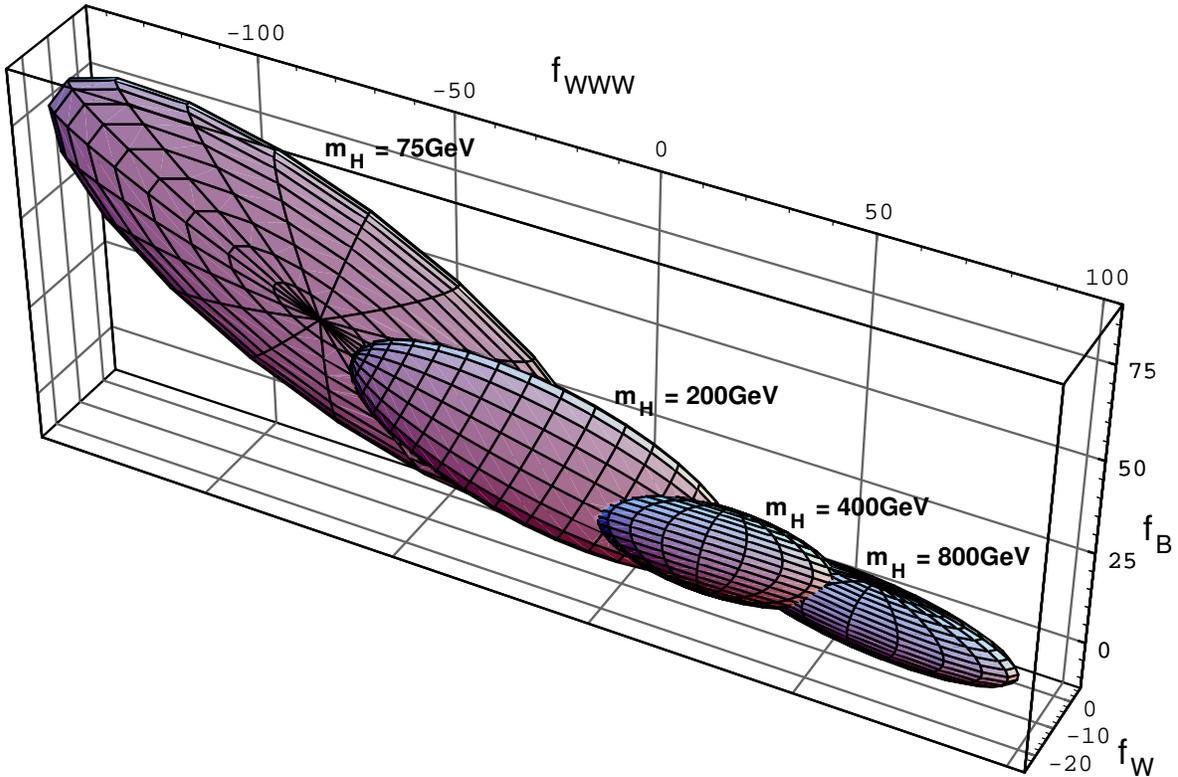,angle=0,width=16cm,silent=0}
\end{center}
\caption{Constraints on $\fwww$, $\fw$ and $\fb$ at the 95\% confidence level 
for $\Lambda = $~1~TeV and $m_t = $~175~GeV.}
\label{fig-fwwwfwfb} 
\end{figure}

For studying non-Abelian gauge-boson self-interactions we are especially 
interested in the operators $\owww$, $\ow$ and $\ob$.  Without presenting an 
explicit model we assume that these are the only relevant couplings and that 
the couplings with tree-level contributions are suppressed.  The results are 
summarised by Fig.~\ref{fig-fwwwfwfb}.
For a light $m_H = $~75~GeV Higgs boson the constraints are rather weak; the 
graphs which contain propagating Higgs bosons tend to cancel against the 
remaining graphs yielding a rather large contour.  The ellipsoid displays 
a strong correlation (anticorrelation) in the $\fwww$--$\fw$ ($\fwww$--$\fb$)
plane.  Notice also that this scenario prefers rather large deviations from the 
SM;  the center of the ellipsoid is at $(\fwww,\fw,\fb) = (-74, -9, 40)$.  As we 
increase $m_H$ the contour becomes smaller and less eccentric, especially for 
$m_H = $~200~GeV or $m_H = $~400~GeV. The $m_H = $~800~GeV contour shows 
flattening in the $\fwww$--$\fw$ plane.  The $m_H = $~200~GeV and 
$m_H = $~400~GeV contours are consistent with the SM while the $m_H = $~75~GeV 
and $m_H = $~800~GeV contours are disfavored.  

\begin{figure}[htb]
\begin{center}
\leavevmode\psfig{file=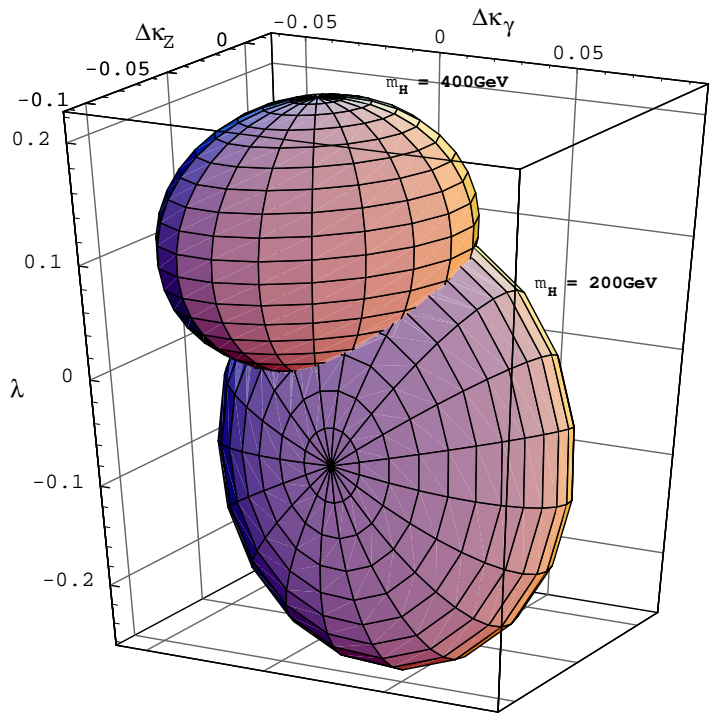,angle=0,width=10cm,silent=0}
\end{center}
\caption{Constraints on $\Delta\kapgam$, $\Delta\kapz$ and $\lambda = \lamgam
= \lamz$ at the 95\% confidence level assuming the relations of 
Eqns.~\protect{\ref{hisz-relations}} with $\Lambda = $~1~TeV and $m_t = $~175~GeV.}
\label{fig-hisz-lin} 
\end{figure}

\begin{figure}[htb]
\begin{center}
\leavevmode\psfig{file=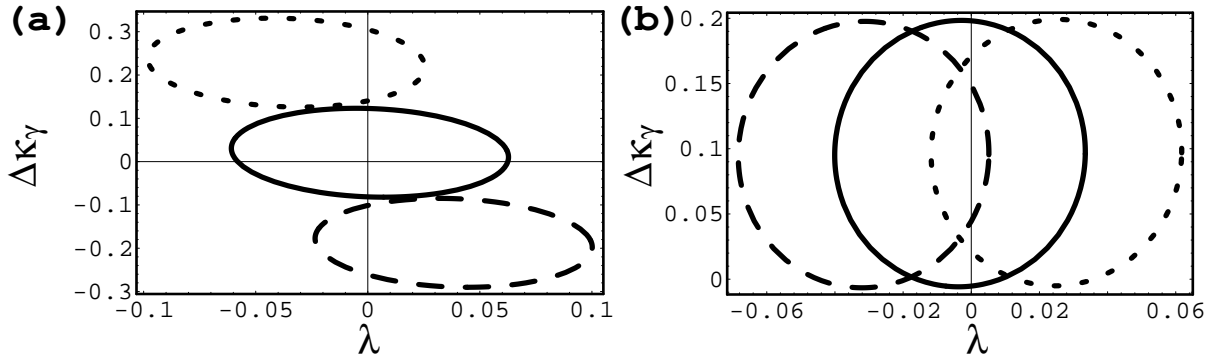,angle=0,width=6.5in,silent=0}
\end{center}
\caption{Constraints in the  $\Delta\kapgam$--$\lambda$ plane at the 
95\% confidence level assuming the relations of 
Eqns.~\protect{\ref{hisz-relations}} with $\Lambda = $~1~TeV and 
$m_t = $~175~GeV.  Fig.~\protect{\ref{fig-hisz-fermilab}}(a) corresponds to 
$m_H = $~200~GeV while Fig.~\protect{\ref{fig-hisz-fermilab}}(b) corresponds to 
$m_H = $~400~GeV.  The solid, dashed and dotted contours correspond to 
$\Delta\kappa_Z = 0$, $-0.1$ and $0.1$ respectively.}
\label{fig-hisz-fermilab} 
\end{figure}

Recall that, by Eqns.~\ref{hisz-relations}, $\fwww$, $\fw$ and $\fb$ are related 
to the standard parameters for nonstandard $WW\gamma$ and $WWZ$ couplings.  The 
95\%~confidence-level contours treating $\Delta\kapgam$, $\Delta\kapz$ and 
$\lambda = \lamgam = \lamz$ as the free parameters are presented in 
Fig.~\ref{fig-hisz-lin} for $m_H = $~200~GeV and $m_H = $~400~GeV. 
Both contours are consistent with the SM, though the $m_H = $~200~GeV contour 
just barely includes the SM value of $\lambda = 0$.  For $m_H = $~200~GeV we 
observe a 
strong $\Delta\kapz$--$\lambda$ correlation which is important when considering 
the measurement of these couplings at the Fermilab Tevatron.  The Tevatron 
is sensitive to the $WW\gamma$ vertex primarily through the observation of
$W\gamma$ pairs, but due to a limited center-of-mass energy $WW$ and $WZ$
events are rare.  Therefore, at the Tevatron we are primarily interested in 
a two-dimensional plot in the $\Delta\kapgam$--$\lambda$ plane with a fixed 
value of $\Delta\kapz$.  Fig.~\ref{fig-hisz-fermilab}(a) is a fit in the 
$\Delta\kapgam$--$\lambda$ plane for $m_H = $~200~GeV, and 
Fig.~\ref{fig-hisz-fermilab}(b) is the same plot for $m_H = $~400~GeV.
The solid, dashed and dotted contours correspond to $\Delta\kappa_Z = 0$, $-0.1$ 
and $0.1$ respectively.  
Notice that the $m_H = $~200~GeV contour with $\Delta\kappa_Z = 0$ is very 
consistent with the SM while all of the $m_H = $~400~GeV contours barely 
cross the $\lambda = 0$ axis.  Clearly there is much more sensitivity to 
the assumed value of $\Delta\kappa_Z$ when the Higgs boson is light.

\subsection{Results for the nonlinear model}

In order to perform the fits in the nonlinear case we calculate the SM values 
using $m_H=1$~TeV.  Correspondingly we take $\hat{\mu}=1$~TeV in 
Eqns.~(\ref{deltas-nl}) and (\ref{deltat-nl}),
 effectively subtracting off the 
Higgs-boson contributions.  This method of subtracting off the SM Higgs-boson 
contributions is approximate, and in principle we should also subtract off all 
of the small finite $m_H$-dependent terms, or we should repeat the calculation 
of the higher-order effects excluding the Higgs boson from the beginning.  
The nonzero expressions for $\rgg$, $\rgz$, $\rzz$ and $\rww$ in 
Eqns.~(\ref{results-nl}) is a clear signal that the one-loop calculations
including \oefour operators have induced \oesix effects.  Therefore, to be 
completely consistent through \oesix, we should add to Eqns.~(\ref{results-nl}) 
the two-loop contributions of the \oetwo operator $\call_1^\prime$ and the 
tree-level contributions of a complete set of \oesix chiral operators.  Excluding
these effects is an approximation, and in order to proceed we must assume that 
the excluded effects do not significantly interfere with the contributions
of $\call_1^\prime$-$\call_{10}$.

First we analyze the numerical constraints on $\alpha_1$, $\beta_1$ and 
$\alpha_8$ (which correspond to $\Delta S$, $\Delta T$ and $\Delta U$, 
respectively), and we present the best-fit central values with one-sigma errors,
\begin{equation}
\label{fit-now-nl}
\left.\begin{array}{lll}
\alpha_1 & = & (4.7 \pm 2.6)\times 10^{-3} \;,\\
\beta_1  & = & (0.30 \pm 0.57)\times 10^{-3}\;,\\
\alpha_8 & = & (-0.9 \pm 7.6)\times 10^{-3} \;,
\end{array}\right.
\end{equation}
and the full correlation matrix,
\begin{equation}
\label{fit-now-nl-corr}
\rho_{\rm corr} = 
\left(\makebox[-0.3cm]{}
\begin{array}{ldd}
\dec 1. & \dec $-$0.871 & \dec $-$0.121 \\
        & \dec    1.    & \dec    0.221 \\
        &               & \dec    1.
\end{array}\right)\;.
\end{equation}
These tree-level contributions are nondecoupling effects, hence the bounds 
derived are insensitive to the scale $\Lambda$.
These constraints are sufficiently strong that there is no sensitivity to 
these three parameters at LEP~II\cite{hms95,hhis96}.  Observe that a positive 
value for $\alpha_1$ is favored.  If we insist that either $\alpha_1 = 0$ or 
$\beta_1 = 0$, then the $\alpha_1$--$\beta_1$ anti-correlation forces the other
parameter towards a more positive central value.  Accordingly, in 
Table~\ref{table-nl-singly-tree}, we present the 95\% confidence-level limits 
where only one of $\alpha_1$, $\beta_1$ or $\alpha_8$ is allowed to deviate from 
zero.
\begin{table}[tbh]
\begin{tabular}{||c|c|c||}
 $\beta_1$ & $\alpha_1$ & $\alpha_8$ 
\\ \hline\hline
 (1.2$\pm$1.0)$\times 10^{-3}$ & (6.0$\pm$4.9)$\times 10^{-3}$ & 
(-7$\pm$28)$\times 10^{-3}$ \\
\end{tabular}
\caption{95\% confidence-level constraints on $\beta_1$, $\alpha_1$ and 
$\alpha_8$ for $m_t = 175$~GeV.  These results are independent of $\Lambda$. In 
this table only one coupling at a time is allowed to deviate from zero.}
\label{table-nl-singly-tree}
\end{table}
Indeed we see that a more positive value for both $\alpha_1$ and $\beta_1$ is 
preferred, and the fitted value of $\beta_1$ now deviates significantly from the 
SM.

Next we place constraints on the remaining parameters by considering the
effects of only one operator at a time.  The results are summarized in 
Table~\ref{table-nl-singly-loop}.
\begin{table}[tbh]
\begin{tabular}{||l||c|c|c|c||}
 & $\alpha_2$ & $\alpha_3$ & $\alpha_9$ &
$\frac{2}{5}\alpha_4 + \alpha_5 + 14.9\alpha_6 + 15.6\alpha_7 
+ 14.7\alpha_{10}$\\
\hline\hline
$\alpha_1 = 0$ & 0.25$\pm$0.20 & -0.12$\pm$0.09 & -0.27$\pm$0.61 
& -0.44$\pm$0.35 \\\hline
$\alpha_1 = 5.5\times 10^{-3}$ & 0.03$\pm$0.20 & -0.05$\pm$0.09 & -0.28$\pm$0.61 
& -0.09$\pm$0.37 \\
\end{tabular}
\caption{95\% confidence-level constraints on $\alpha_2$--$\alpha_7$, $\alpha_9$
and $\alpha_{10}$ for $m_t = 175$~GeV and $\Lambda = 1$~TeV.  Only the linear 
combination of $\alpha_4$--$\alpha_7$ and $\alpha_{10}$ shown in the last column
may be constrained.  In the first row all other coefficients are set to zero.  In 
the second row $\alpha_1 = 5.5\times 10^{-3}$ is chosen according to 
Eqn.(\protect{\ref{fit-now-nl}}).}
\label{table-nl-singly-loop}
\end{table}
First of all, notice that $\alpha_4$--$\alpha_7$ and $\alpha_{10}$ enter into 
the analysis only through their contribution to $\Delta T$ as shown in 
Eqn.~(\ref{deltat-nl}), hence only the linear combination of these five 
coefficients shown in the last column may be constrained.  Furthermore, 
$\alpha_1$ is anticorrelated with this linear combination in the same fashion as 
with $\beta_1$.  Notice that in the first row of the table, when $\alpha_1 = 0$,
only $\alpha_9$ is consistent with SM at the 95\% confidence level.  However,
in the second row where we have chosen the central value of $\alpha_1$ according 
to the best-fit value of Eqn.~(\ref{fit-now-nl}), all of the central values 
are easily consistent with the zero.  While the central values easily move 
around as we include additional operators in the analysis, the errors are 
much more robust.

\begin{figure}[htb]
\begin{center}
\leavevmode\psfig{file=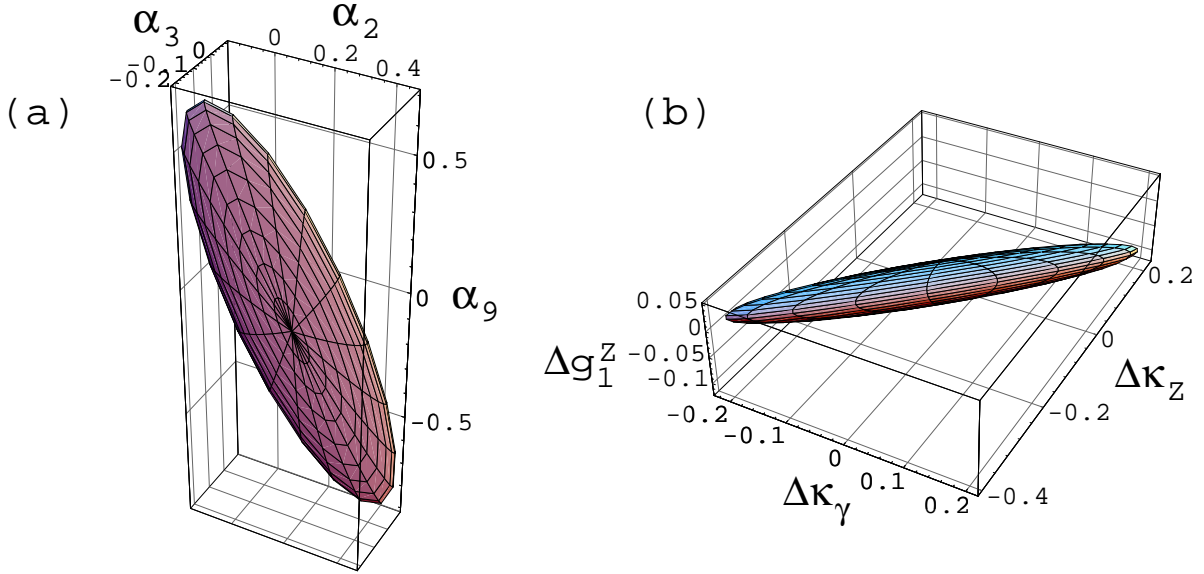,angle=0,width=6.5in,silent=0}
\end{center}
\caption{95\% confidence level contours for $\Lambda = $~1~TeV and 
$m_t = $~175~GeV.  In (a) we show the fit in $\alpha_2$, $\alpha_3$ and 
$\alpha_9$.  In (b) we have reparameterized the fit in terms of $\Delta\kapgam$,
$\Delta\kapz$ and $\Delta g_1^Z$ according to the relations of 
Eqns.~(\protect{\ref{nl-hisz-relations}}). }
\label{fig-hisz-nl} 
\end{figure}

Three of the coefficients, $\alpha_2$, $\alpha_3$ and $\alpha_9$, contribute
at the tree-level to nonstandard $WW\gamma$ and $WWZ$ vertices without 
making a tree-level contribution to low-energy and Z-pole observables.  In 
Fig.~\ref{fig-hisz-nl}(a) we plot 95\% confidence-level limits obtained
by fitting $\alpha_2$, $\alpha_3$ and $\alpha_9$.  There is a very strong 
$\alpha_2$--$\alpha_3$ correlation and moderately strong $\alpha_2$--$\alpha_9$
and $\alpha_3$--$\alpha_9$ anti-correlations.  Then, using 
Eqns.~(\ref{nl-hisz-relations}), we may recast the fit in terms of 
$\Delta\kapgam$, $\Delta\kapz$ and $\Delta g_1^Z$.  The results are displayed 
in Fig.~\ref{fig-hisz-nl}(b).  In this basis the correlations are not as strong;
there are moderately strong $\Delta\kapgam$--$\Delta\kapz$ and 
$\Delta\kapgam$--$\Delta g_1^Z$ correlations.  In Fig.~\ref{fig-hisz-nl}(a)
the point $\alpha_3 = 0$ (equivalently, in Fig.~\ref{fig-hisz-nl}(b), 
the point $\Delta g_1^Z = 0$) lies near the edge of the contour.

\begin{figure}[htb]
\begin{center}
\leavevmode\psfig{file=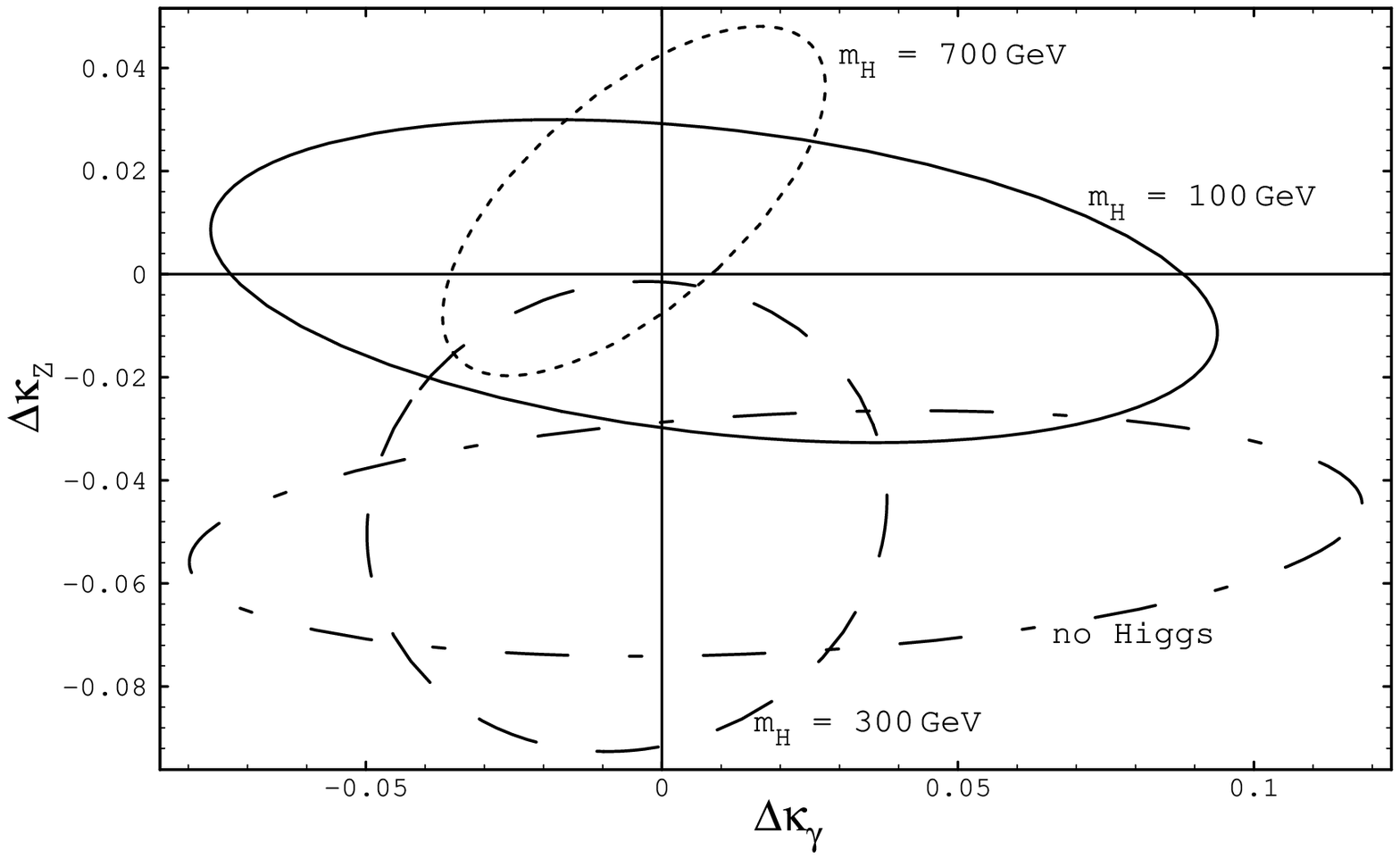,angle=0,width=5in,silent=0}
\end{center}
\caption{95\% confidence level contours for $\Lambda = $~1~TeV and 
$m_t = $~175~GeV.  The solid, dashed and dotted curves correspond to 
$m_H = 100$~GeV, $m_H = 300$~GeV and  $m_H = 700$~GeV; these first three 
curves use Eqns.~(\protect{\ref{hisz-relations}}).  The dot-dashed curve 
corresponds to the nonlinear realization of SSB and therefore employs 
Eqns.~(\protect{\ref{nl-hisz-relations}}).}
\label{fig-lin-nl} 
\end{figure}

If we require any new physics to conserve the SU(2)$_{\rm C}$ symmetry, then
$\alpha_9 = 0$.  In this case there are only two free parameters, $\alpha_2$ 
and $\alpha_3$; equivalently we can choose any two parameters from the set 
$\{\Delta\kapgam, \Delta\kapz, \Delta g_1^Z\}$, and once again we use 
the relations of Eqns.~(\ref{nl-hisz-relations}).  Recalling that 
$\alpha_2$ and $\alpha_3$ are related to $\fb$ and $\fw$ 
by Eqns.~(\ref{relate4}) we can perform the analogous fit in the 
linear realization of SSB which is shown in 
 Fig.~\ref{fig-lin-nl}.
The solid, dashed and dotted curves correspond to 
$m_H = 100$~GeV, $m_H = 300$~GeV and  $m_H = 700$~GeV; these first three
curves use Eqns.~(\ref{hisz-relations}).  The $m_H = 100$~GeV is very 
consistent with the SM while the $m_H = 300$~GeV and $m_H = 700$~GeV contours 
prefer nonzero values for $\Delta\kapz$; the centers and the orientations of 
these ellipses are complicated functions of $m_H$, but the contours clearly 
become smaller with increasing Higgs-boson mass. The dot-dashed curve 
corresponds to the nonlinear realization of SSB and therefore employs 
Eqns.~(\ref{nl-hisz-relations}).  It clearly does not include the SM, but its 
center could be shifted by including nonzero central values for $\alpha_1$ and 
$\beta_1$ according to Eqn.~(\ref{fit-now-nl}).

In any realistic scenario there will be a set of nonzero $\alpha_i$, and it is 
possible (indeed likely) that there will be large interference between the 
effects of the various coefficients.  In order to see the types of limits which 
might arise in various scenarios of SSB we consider a strongly interacting 
scalar and a degenerate doublet of heavy fermions, and we get an indication of 
the sensitivity  of our results to the underlying dynamics.  Using the 
effective-Lagrangian approach, we can estimate the coefficients in a consistent 
way.

We first consider a model with three Goldstone bosons corresponding to the 
longitudinal components of the $W^\pm$ and $Z$ and bosons coupled to a scalar 
isoscalar resonance like the Higgs boson.  We assume that the $\alpha_i(\mu^2)$ 
are dominated by tree-level exchange of the scalar boson.  Integrating out the 
scalar and matching the coefficients at the scale $m_H$ gives the 
predictions\cite{ab80,drv89,hm95},
\begin{mathletters}
\label{scalar-alphai}
\begin{eqnarray}
\label{scalar-alpha4}
\alpha_4(\mu^2) & = & 
   \frac{1}{16\pi^2}\frac{1}{12} \log\bigg(\frac{\mhsq}{\mu^2}\bigg)
\\ & = & 2\alpha_2(\mu^2)=2\alpha_3(\mu^2)=-\alpha_1(\mu^2)
\label{scalar-alpha321}
\\
\alpha_5(\mu^2) & = & \frac{1}{16\pi^2}\Bigg[\frac{1}{24} 
\log\bigg(\frac{\mhsq}{\mu^2}\bigg)
+\frac{64\pi^3}{3}\frac{\Gamma_h v^4}{m_H^5}\Bigg],
\label{scalar-alpha5}
\end{eqnarray}
\end{mathletters}
where $\Gamma_h$ is the width of the scalar into Goldstone bosons.  All of the 
other $\alpha_i$ are zero in this scenario.  It is important to note that only 
the logarithmic terms are uniquely specified.  The constant terms depend on the 
renormalization scheme\cite{hm95,bdv93}. (We use the renormalization scheme of 
Ref.~\cite{bdv93}.)   

\begin{figure}[htb]
\begin{center}
\leavevmode\psfig{file=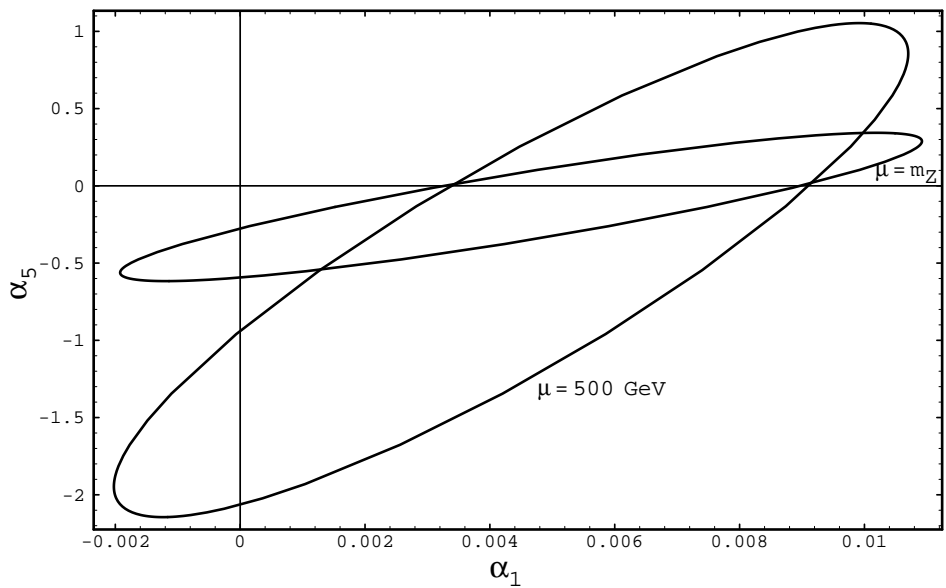,angle=0,width=5in,silent=0}
\end{center}
\caption{95\% confidence level contours for $\Lambda = $~1~TeV and 
$m_t = $~175~GeV in the $\alpha_5$--$\alpha_1$ plane subject to 
$-\alpha_1(\mu^2) = 2\alpha_2(\mu^2) = 2\alpha_3(\mu^2) = \alpha_4(\mu^2)$.
The larger (smaller) contour corresponds to $\mu = 500$~GeV ($\mu = m_Z$).}
\label{fig-scalar-model} 
\end{figure}

In Fig.~\ref{fig-scalar-model}
we plot $\alpha_5(\mu^2)$ 
{\em vs.}\/ $\alpha_1(\mu^2)$ with the pattern typical of a theory dominated by 
a heavy scalar given in Eqn.~(\ref{scalar-alphai}), 
$-\alpha_1(\mu^2) = 2\alpha_2(\mu^2) = 2\alpha_3(\mu^2) = \alpha_4(\mu^2)$.  
First of all, notice that the contour obtained depends rather strongly upon 
our choice of the renormalization scale, $\mu$, especially with regard to the 
$\alpha_5$ axis. Everything to the right of $\alpha_1 = 0$ corresponds to 
$m_H < \mu$.  Furthermore, since we require that $\Gamma_h$ be non-negative,
we may approximately exclude everything below the $\alpha_5 = 0$ axis.  The 
allowed region to the upper right of the figure corresponds to a Higgs-boson 
with a mass in the MeV range and an extremely narrow width; this portion of 
the figure is already excluded by experiment.  An approximate upper bound on 
$m_H$ can be obtained from the leftmost point where both curves intersect 
the horizontal axis; as this figure is drawn the entire plane is excluded by 
LEP.  We can, by changing $\Lambda$ and $\mu$, drive the upper bound above 
100~GeV or greater, but the positive central value of $\alpha_1$ indicates that 
a heavy scalar resonance is disfavored.

The previous example conserves the custodial SU(2)$_{\rm C}$ symmetry.  The 
simplest 
example of dynamics which violates the custodial symmetry is a heavy doublet of 
nondegenerate fermions. Ref.~\cite{aw93} considers the case of a heavy doublet 
with charge $\pm\frac{1}{2}$, a mass splitting $\Delta m$, and an average mass 
$M$ with ($\Delta m \ll M$).  Then assuming the fermions are  in a color triplet
and retaining terms to ${\cal O}(\delta^2)$, ($\delta \equiv 
{\Delta m\over 2M}$), they find
\begin{mathletters}
\label{fermalph}
\begin{eqnarray}
\label{fermalph-b1}
\beta_1(\mzsq) & = & \frac{1}{8\pi^2}\frac{(\Delta m)^2}{m_W^2} \;,
\\
\label{fermalph-a1-a2}
\alpha_1(\mzsq) & = & \alpha_2(\mzsq) = -\frac{1}{32\pi^2} \;,
\\
\label{fermalph-a3}
\alpha_3(\mzsq) & = & -\frac{1}{32\pi^2}\bigg(1-\frac{2}{5}\delta^2 \bigg) \;,
\\
\label{fermalph-a8}
\alpha_8(\mzsq) & = & -\frac{1}{32\pi^2}\frac{16}{5}\delta^2 \;,
\\
\label{fermalph-a9}
\alpha_9(\mzsq) & = & -\frac{1}{32\pi^2}\frac{14}{5}\delta^2 \;,
\\
\label{fermalph-a11}
\alpha_{11}(\mzsq) & = & \frac{1}{32\pi^2}\delta \;,
\end{eqnarray}
\end{mathletters}
Because of the heavy fermion masses in the loops, the $\alpha_i$ are finite and 
there are no logarithms of $\Lambda$ in Eqns.~(\ref{fermalph}).  The custodial
SU(2)$_{\rm C}$ violation can be clearly seen in the terms proportional to 
$\delta$.  As in the case of the heavy Higgs boson, we note that the 
coefficients are naturally ${\cal O}(1/16\pi^2)$.  (For a discussion where the 
mass splitting is arbitrary, see Ref.~\cite{fmrs95}.) 

\begin{figure}[tb]
\begin{center}
\leavevmode\psfig{file=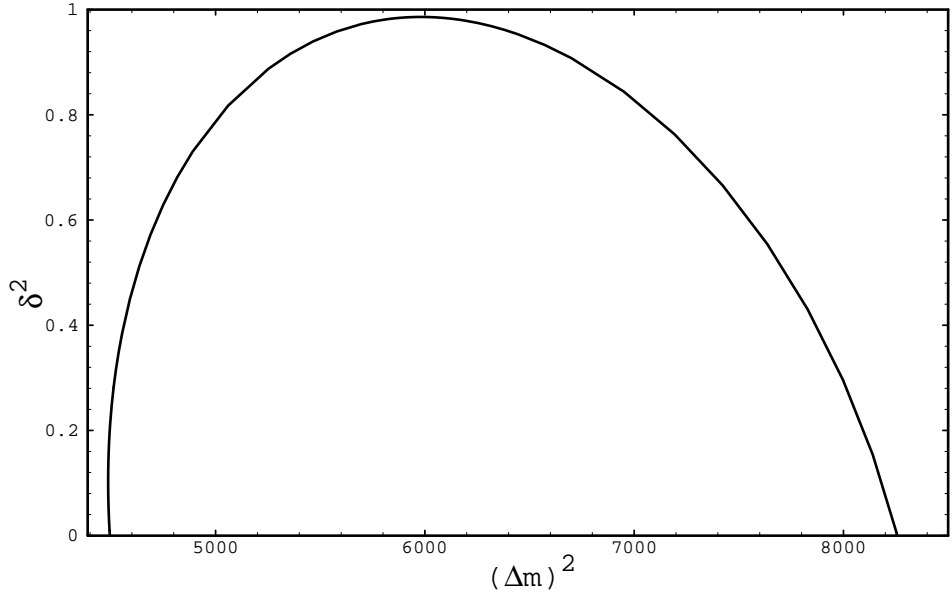,angle=0,width=5in,silent=0}
\end{center}
\caption{95\% confidence level contours for in the $(\Delta m)^2$--$\delta^2$ 
using the parameters of Eqn.~{\protect{\ref{fermalph}}}.  Numerically there is 
an allowed region for $\delta^2<0$.  However, $\delta^2<0$ is unphysical, and 
we do not show it.  All numbers are in GeV$^2$.}
\label{fig-aw} 
\end{figure}

This model generates a nonzero value for $\alpha_{11}$, but we have not 
included $\alpha_{11}$ in our analysis.  This is not a problem since we 
expect the analysis to be dominated by the tree-level contributions of 
$\beta_1$, $\alpha_1$ and $\alpha_8$; we will neglect the contributions of
the other coefficients.  In Fig.~\ref{fig-aw}  we show the 95\% confidence-level 
limits in the $(\Delta m)^2$--$\delta^2$ plane. 
We have excluded the unphysical $\delta^2<0$ portion of the ellipse.  However,
the calculation is not valid for a portion of the region shown.  We have 
explicitly assumed that $\Delta m \ll M$.  If we choose a very loose cut-off 
of $\Delta m < 0.4 M$, then we should restrict the figure to $\delta^2 < 0.04$,
and only a narrow strip along the bottom of the figure is relevant. 
For $\delta^2 = 0$, $M\rightarrow \infty$, and we cannot obtain an upper 
bound on $M$.  We cannot obtain a lower bound on $M$ because the contour extends 
into a region where our calculation is not valid.  If we insist that the new 
fermions are heavier than approximately 200~GeV, then 
$ \Delta M \sim {\cal O}(60-90~{\rm GeV})$ is the preferred region for the 
mass splitting.  


\section{Conclusions}

Parameterizing the contributions of new physics at low energies with an effective
Lagrangian we have studied the contributions of new physics to electroweak 
observables; everywhere we have treated the linear and nonlinear realizations 
of electroweak symmetry breaking in parallel, allowing us to make direct 
comparisons which had not previously been studied.  The complete contributions 
of the new physics to low-energy and Z-pole observables may be completely 
summarised by expressions for the running charges $\baralphaqsq$, $\bargzsqqsq$,
$\barssqqsq$ and $\bargwsqqsq$ plus a form factor for the $Zbb$ vertex, 
$\bardeltab(\qsq)$.  We present explicit expressions for these quantities in 
both realizations of symmetry breaking.

The above approach is ideally suited to performing a global analysis using all
available electroweak precision data.  We perform many such fits.  We study the 
bounds which may be obtained on the various effective-Lagrangian 
parameters and the bounds on nonstandard $WW\gamma$ and $WWZ$ couplings.  For 
the case of nonstandard $WW\gamma$ and $WWZ$ couplings we are able to investigate
the role of the Higgs mass as compared to having no Higgs boson as all.  See
Fig.~\ref{fig-lin-nl}.  

The coefficients of some operators in the effective Lagrangian contribute to 
four-fermion amplitudes at the tree level while the coefficients of others 
first contribute at the loop level.  A topic of great interest is whether the 
former can be suppressed relative to the latter.  We discuss one toy model 
where such a hierarchy is realized.  If such a hierarchy could be realized 
among the operators that contribute to $WW\gamma$ and $WWZ$ couplings, then,
even allowing for some correlations, the low-energy bounds are in some 
cases on par with or even superior to the bounds that can be obtained at 
LEP~II.

We then use our global analysis to examine some explicit models.  For the case 
of a strongly interacting model with a scalar Higgs boson, a light scalar is 
strongly prefered, while much of the light region has already been ruled out 
by LEP.  We confirm that a positive value for $\alpha_1$ is prefered, which 
is known to strongly disfavor the simplest models that include a strongly 
interacting vector-like Higgs boson.  Finally we consider the contributions
of a heavy pair of new fermions.  While our analysis is only valid if their 
masses are heavier than 200-300~GeV, we find that a mass splitting of 60-90~GeV
is prefered.


\section*{Acknowledgements}
Special thanks to Seiji Matsumoto for prior collaboration on related 
works and for providing us with an updated analysis of the electroweak 
data.  We are grateful to Cliff Burgess and Dieter Zeppenfeld for 
stimulating discussions.  The contributions of Rob Szalapski were 
supported in part by the National Science Foundation through grant no.
INT9600243 and in part by the Japan Society for the Promotion of Science
(JSPS).  The work of S. Dawson supported by U.S. Department of Energy
under contract DE-AC02-76CH00016.  The work of S. Alam was supported in part by 
a COE Fellowship from the Japanese Ministry of Education and Culture and in 
part by JSPS.

\appendix
\section{Operators in the linear realization of SSB}
\label{app-linear}

In this appendix we explicitly enumerate the operators of Eqn.~(\ref{leff-lr}),
{\em i.e.\/} the effective Lagrangian with the linear realization of SSB.  The 
twelve operators discussed in Section~\ref{sec-linear} are 
\begin{mathletters}
\label{twelve}
\begin{eqnarray}
\label{odw}
{\cal O}_{DW} & = & - \hat{g}^2 \trace 
  \bigg[ \Big(\partial_\mu W_{\nu\rho} \Big) 
 \Big( \partial^\mu W^{\nu\rho}\Big)\bigg] \;,\\
\label{odb}
{\cal O}_{DB} & = & -\frac{\hat{g}^{\prime\;2}}{2}\Big(\partial_\mu
B_{\nu\rho}\Big)\Big(\partial^\mu B^{\nu\rho}\Big) \;,\\
\label{obw}
{\cal O}_{BW} & = & - \frac{\hat{g} \hat{g}^\prime}{2} 
   \Phi^\dagger B_{\mu\nu}W^{\mu\nu} \Phi  \;,\\
\label{ophi1}
{\cal O}_{\Phi,1} & = & \bigg[ \Big(D_\mu\Phi\Big)^\dagger \Phi\bigg] \; 
\bigg[ \Phi^\dagger \Big(D^\mu \Phi\Big)\bigg]\;,\\
\label{owww}
{\cal O}_{WWW} & = & -i\frac{3}{2}\hat{g}^3 W^+_{\mu\nu}W^{-\;\nu\rho}
W^{3\;\mu}_{\rho}\;,\\
\label{ow}
{\cal O}_{W} & = & i\hat{g}\Big(D_\mu\Phi\Big)^\dagger 
                   W^{\mu\nu}\Big(D_\nu\Phi\Big)\;,\\
\label{ob}
{\cal O}_{B} & = & \frac{i}{2}\hat{g}^{\prime}
\Big(D_\mu\Phi\Big)^\dagger  B^{\mu\nu}
     \Big(D_\nu\Phi\Big)\;,\\
\label{oww}
{\cal O}_{WW} & = & -\frac{\hat{g}^2}{4}\Big(\Phi^\dagger \Phi\Big)
W^{I\,\mu\nu}W^I_{\mu\nu}  \;,\\
\label{obb}
{\cal O}_{BB} & = & -\frac{\hat{g}^{\prime\,2}}{4}\Big(\Phi^\dagger \Phi\Big)
B^{\mu\nu}B_{\mu\nu}  \;,\\
\label{ophi2}
{\cal O}_{\Phi,2} & = & \frac{1}{2}\partial_\mu\Big(\Phi^\dagger \Phi\Big) 
\partial^\mu\Big(\Phi^\dagger \Phi\Big)  \;,\\
\label{ophi3}
{\cal O}_{\Phi,3} & = & \frac{1}{3}\Big(\Phi^\dagger \Phi\Big)^3 \;,\\
\label{ophi4}
{\cal O}_{\Phi,4} & = & \Big(\Phi^\dagger \Phi\Big)\bigg[
\Big(D_\mu \Phi\Big)^\dagger \Big(D^\mu \Phi\Big) \bigg]  \;,
\end{eqnarray}
\end{mathletters}
where $W_{\mu\nu} = T^a W^{a}_{\mu\nu}$.  The field strength tensors are 
given by
\begin{mathletters}
\label{field-strenths}
\begin{eqnarray}
\label{bmunu}
B_{\mu\nu} & = & \partial_\mu B_\nu - \partial_\nu B_\mu  \;,\\
\label{wmunu}
W_{\mu\nu} & = & \partial_\mu W_\nu - \partial_\nu W_\mu - g\epsilon_{abc}
W_\mu^b W_\nu^c \;,
\end{eqnarray}
\end{mathletters}
where $\epsilon_{abc}$ is the totally antisymmetric tensor in three dimensions
with $\epsilon_{123} = 1$.  The covariant derivative is given by 
\begin{equation}
D_\mu = \partial_\mu + i g T^a W^a_\mu + i g^\prime Y B_\mu,
\end{equation}
and $\Phi$ is the SM Higgs doublet,
\begin{equation}
\Phi = \frac{1}{\sqrt{2}}\left( 
\begin{array}{c} 
i\chi^1+\chi^2 \\ 
v + H - i \chi^3 
\end{array} 
\right)\;.
\end{equation}

\section{Operators in the electroweak chiral Lagrangian}
\label{app-nonlinear}

In this appendix we present explicitly the operators of the electroweak 
chiral Lagrangian.  In the notation of Ref.~\cite{lon80,lon81,aw93}, 
\begin{equation}
\label{leff-nlr2}
\call_{\rm eff}^{\rm nlr} = \call_{\rm SM}^{\rm nlr} + 
\sum \call_{i} + \cdots \;.
\end{equation}
We use the superscript `nlr', denoting `nonlinear realization'.  The 
first term is the SM Lagrangian, but in this case no physical Higgs boson
is included.  Hence $\call_{\rm SM}^{\rm nlr}$  is nonrenormalizable.
The first non-SM terms are energy dimension-two and -four 
operators which are not manifestly suppressed by explicit powers of some high 
scale.  

While the physical Higgs boson has not been employed, the Goldstone 
bosons, $\chi_i$ for $i = 1,2,3$, are included  through the unitary unimodular 
field $U$ introduced below.  Following Ref.~\cite{aw93}, 
\begin{mathletters}
\begin{eqnarray}
U & \equiv & 
\exp \bigg( \frac{2i\,\chi_i(x)\tau_i}{v}\bigg)
\longrightarrow {\bf 1} \;, \\
D_\mu U & \equiv & \partial_\mu U + i g T^a W_\mu^a U - i g^\prime U T^3 B_\mu 
\longrightarrow i g T^a W_\mu^a - i g^\prime T^3 B_\mu \;, \\
\label{tee-general}
T & \equiv & 2 U T^3 U^\dagger \longrightarrow 2 T^3 \;, \\
V_\mu & \equiv & ( D_\mu U ) U^\dagger 
\longrightarrow D_\mu U \;,
\end{eqnarray}
\end{mathletters}
where $\bf 1$ is the 2$\times$2 identity matrix, the Pauli matrices are denoted 
by $\tau_i$, and $T_i = \tau_i/2$ with the normalization 
Tr$(T_iT_j) = \frac{1}{2}
\delta_{ij}$.  The right-pointing arrow indicates the unitary-gauge form of each 
expression.        The lowest order effective Lagrangian for the
symmetry breaking sector of the theory is
\begin{equation}
\call_{SM}={v^2\over 4} Tr\biggl[
D_\mu U^\dagger D^\mu U\biggr]
\quad .
\end{equation}

The non-SM operators with four or fewer derivatives which conserve 
CP are\cite{lon81,aw93},
\begin{mathletters}
\label{twelve-chiral}
\begin{eqnarray}
\label{l1p}
\call_1^\prime 
  & = & \frac{1}{4}\beta_1 v^2 \Big[ \trace ( T V_\mu ) \Big]^2 \;,\\
\label{l1}
\call_1 & = & \frac{1}{2}\alpha_1 \hatg\hatg^\prime\trace 
                 \Big(B_{\mu\nu} T W^{\mu\nu} \Big) \;,\\
\label{l2}
\call_2 & = & \frac{i}{2} \alpha_2 \hatg^\prime
        B_{\mu\nu} \trace \Big( T [ V^\mu , V^\nu ] \Big) \;,\\
\label{l3}
\call_3 & = & i \alpha_3 \hatg\trace \Big( W_{\mu\nu} [V^\mu , V^\nu ] \Big) \;,\\
\label{l4}
\call_4 & = & \alpha_4 \Big[ \trace ( V_\mu V_\nu ) \Big]^2 \;,\\
\label{l5}
\call_5 & = & \alpha_5 \Big[ \trace ( V_\mu  V^\mu ) \Big]^2 \;,\\
\label{l6}
\call_6 & = & \alpha_6 \trace \Big( V_\mu V_\nu \Big) 
        \trace \Big( T V^\mu \Big) \trace \Big( T V^\nu \Big) \;,\\
\label{l7}
\call_7 & = & \alpha_7 \trace \Big( V_\mu V^\mu \Big) 
        \trace \Big( T V_\nu \Big) \trace \Big( T V^\nu \Big) \;,\\
\label{l8}
\call_8 & = & \frac{1}{4}\alpha_8 \hatgsq \Big[ \trace(TW_{\mu\nu})\Big]^2 \;,\\
\label{l9}
\call_9 & = & \frac{i}{2}\alpha_9 \hatg \trace \Big( T W_{\mu\nu} \Big) 
        \trace \Big( T [V^\mu , V^\nu ] \Big) \;,\\
\label{l10}
\call_{10} & = & \frac{1}{2} \alpha_{10}
        \Big[ \trace ( T V_\mu ) \trace ( T V_\nu ) \Big]^2 \;,\\
\label{l11}
\call_{11} & = & \alpha_{11} \hatg\, \epsilon^{\mu\nu\rho\sigma}
\trace \Big( T V_\mu \Big) \trace \Big( V_\nu W_{\rho\sigma} \Big) \;.
\end{eqnarray}
\end{mathletters}


\end{document}